\documentclass[a4paper]{report}
\usepackage{amsmath,german}
\usepackage{amsfonts}
\usepackage{pstricks}
\usepackage{bbm}
\usepackage{amstext}
\usepackage{latexsym}
\usepackage[dvips]{graphicx}
\usepackage{psfrag}

\def\DJo{$\;$\kern-.4em \hbox{D\kern-.8em\raise.15ex\hbox{--}\kern.35em okovi\'c}}
\def\be{\begin{eqnarray}}
\def\bea{\begin{eqnarray}}
\def\bma{\begin{mathletters}}
\def\ee{\end{eqnarray}}
\def\eea{\end{eqnarray}}
\def\ema{\end{mathletters}}

\newcommand{\nc}{\newcommand}
\nc{\rnc}{\renewcommand} \nc{\ket}[1]{\left | \, #1 \right
\rangle} \nc{\bra}[1]{\left \langle #1 \, \right |}
\nc{\proj}[1]{\ket{#1}\bra{#1}} \rnc{\vec}{\mathbf}
\nc{\braket}[2]{\langle\, #1\,|\,#2\,\rangle}
\nc{\half}{\frac{1}{2}}
\nc{\vfigure}[3]{
\begin{figure}[th]
\centerline{\psfig{file=figures/#1.eps,width=#2}}
\vspace*{8pt}
\caption{#3}
\end{figure}}
\nc{\vpstexfigure}[3]{
\vfigure{#1}{#2}{#3}}
\nc{\prj}{\mathcal{P}} \nc{\hilb}{\mathcal{H}}
\nc{\pth}{\mathcal{C}} \nc{\inprod}[2]{\braket{#1}{#2}}
\nc{\upket}{\ket{\uparrow}} \nc{\downket}{\ket{\downarrow}}
\nc{\upbra}{\bra{\uparrow}} \nc{\downbra}{\bra{\downarrow}}

\def\CC{{\rm\kern.24em \vrule width.04em height1.46ex depth-.07ex
\kern-.30em C}}
\def\P{{\rm I\kern-.25em P}}
\def\N{{\rm I\kern-.25em N}}
\def\RR{{\rm
         \vrule width.04em height1.58ex depth-.0ex
         \kern-.04em R}}
\def\id{{\rm 1\kern-.26em l}}
\def\ZZ{{\sf Z\kern-.44em Z}}
\def\e{{\rm e}}

\def\eps{\varepsilon}
\def\trace{{\rm tr}\;}

\def\d{{\rm d}}
\def\up{\uparrow}
\def\down{\downarrow}
\newtheorem{lemma}{Lemma}[section]
\newtheorem{theorem}{Theorem}[section]
\newcommand{\Matrix}[2]{\left( \begin{array}{#1} #2 \end{array}
  \right)}
\newcommand{\diag}{{\rm diag}\;}

\newenvironment{eqblock}[2]{\beq\label{#2}\begin{array}{#1}}{\end{array}
                                \eeq}
\newenvironment{neqblock}[1]{\[\begin{array}{#1}}{\end{array}\]}
\newcommand{\fat}[1]{\mbox{\boldmath $ #1 $\unboldmath}}
\newcommand{\beqb}{\begin{eqblock}}
\newcommand{\eeqb}{\end{eqblock}} 
\newcommand{\nbeqb}{\begin{neqblock}}
\newcommand{\neeqb}{\end{neqblock}} 
\newcommand{\bigfrac}[2]{\mbox {${\displaystyle \frac{ #1 }{ #2 }}$}}
\newcommand{\beq}{\begin{equation}}
\newcommand{\beqa}{\begin{eqnarray}}
\newcommand{\eeq}{\end{equation}}
\newcommand{\eeqa}{\end{eqnarray}}
\newcommand{\nbeqa}{\begin{eqnarray*}}
\newcommand{\neeqa}{\end{eqnarray*}}
\newcommand{\Expect}[3]{\bra{#1} #2 \ket{#3}}
\newcommand{\expect}[1]{\left\langle #1 \right\rangle}

\begin{document}

\begin{center}
{\Huge\bf Entanglement and its facets in~condensed~matter~systems}

\vspace*{17mm}

{\Large\bf Andreas Osterloh}

\vspace*{34mm}

\hspace*{17mm}\includegraphics[width=\textwidth]{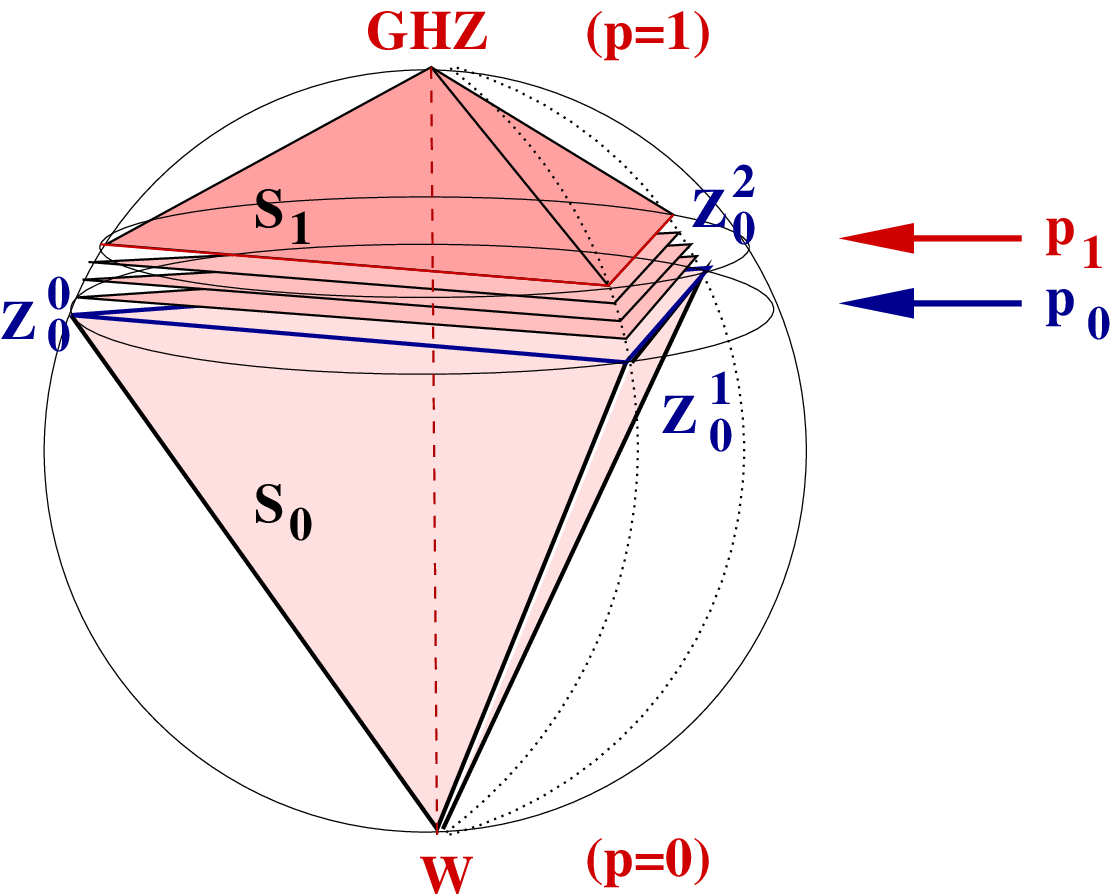}

\vfill
\noindent{\large\bf Der Fakult\"at f\"ur 
Mathematik und Physik der \\
Gottfried Wilhelm Leibniz Universit\"at Hannover\\
zur Habilitation vorgelegte wissenschaftliche Abhandlung}
\end{center}

\thispagestyle{empty}

\newpage

${}$
\thispagestyle{empty}

\newpage

\thispagestyle{empty}

\begin{center}
{\huge \bf \vspace*{-22mm}Erkl\"arung}
\end{center}
{\large
Die vorliegende Habilitationsschrift gibt eine \"Ubersicht
\"uber Forschungsergebnisse, welche bereits - bis auf Paragraph~\ref{Exp} -
in wissenschaftlichen Zeitschriften ver\"offentlicht wurden.
Die relevanten Publikationen dazu sind}
\begin{itemize}
\item \parbox[t]{11cm}{A. Osterloh, L. Amico, G. Falci, and R. Fazio,
	{\em Scaling of the Entanglement close to 
	Quantum Phase Transitions}, Nature {\bf 416}, 608-610 (2002).}

\item \parbox[t]{11cm}{L. Amico and A. Osterloh, 
	{\em Out of equilibrium correlation functions of 
	quantum anisotropic XY models: one-particle excitations}, 
	J.Phys. A {\bf 37}, 291 (2004). }
 
\item \parbox[t]{11cm}{L. Amico, A. Osterloh, F. Plastina, 
	R. Fazio, and M. Palma,
	{\em Dynamics of Entanglement in One-Dimensional Spin Systems}, 
	Phys. Rev. A. {\bf 69}, 022304 (2004).}\\
\parbox[t]{11cm}{\small Selected for publication in the February issue of the 
{\em Virtual Journal of Nanoscale Science \& Technology}, 2004 and \\
in the February issue of the {\em Virtual Journal of Quantum Information}, 2004.}

\item \parbox[t]{11cm}{F. Plastina, L. Amico, A. Osterloh, and 
        R. Fazio, 
	{\em Spin wave contribution to entanglement in Heisenberg models},
	New J. Phys. {\bf 6}, 124 (2004).}

\item \parbox[t]{11cm}{L. Amico, A. Osterloh, F. Plastina, and
	R. Fazio,
	{\em Entanglement in One-Dimensional Spin Systems},
	to appear in the Proceedings of the SPIE {\em defense and security},
	Orlando, Florida, USA (2004).}

\item \parbox[t]{11cm}{A. Osterloh, J. Siewert,
	{\em Constructing N-qubit entanglement monotones from anti-linear 
	  operators}, 
	Phys. Rev. A {\bf 72}, 012337 (2005).}\\ 
\parbox[t]{11cm}{\small Selected for publication in the
August 2005 issue of the {\em Virtual Journal of Quantum Information}}.

\item \parbox[t]{11cm}{A. Osterloh and J. Siewert,
	{\em Entanglement monotones and maximally entangled 
	  states for multipartite qubit systems},
	Int. J. Quant. Inf. {\bf 4}, 531 (2006).}

\item \parbox[t]{11cm}{R. Lohmeyer, A. Osterloh, J. Siewert, 
        and A. Uhlmann,
       {\em Entangled three-qubit states without concurrence
           and three-tangle},
        Phys. Rev. Lett {\bf 97}, 260502 (2006).}

\item \parbox[t]{11cm}{A. Osterloh, G. Palacios, and S. Montangero
       {\em Enhancement of pairwise entanglement from $\mathbbm{Z}_2$ 
        symmetry breaking}, 
       Phys. Rev. Lett {\bf 97}, 257201 (2006).}\\ 
\parbox[t]{11cm}{\small Selected for the July 3, 2007 issue of the
{\em Virtual Journal of Nanoscale Science \& Technology}.}

\item \parbox[t]{11cm}{L. Amico, R. Fazio, A. Osterloh, and V. Vedral,
       review on {\em Entanglement in Many-Body Systems},
       quant-ph/0703044, submitted to Review of Modern Physics.}

\end{itemize}

{\large\noindent
Ich erkl\"are hiermit, da\3 alle zur Ausf\"uhrung der vorliegenden Arbeit
ben\"otigten Hilfsmittel und Referenzen zitiert werden.

\vspace*{34mm}
\hspace*{17mm}{\large Andreas Osterloh}}
  
\newpage

${}$
\newpage

\thispagestyle{empty}
\begin{center}
{\huge \bf Danksagung}
\end{center}

\vspace*{3mm}
{\large
An dieser Stelle m\"ochte ich all denjenigen danken, die das Entstehen
dieser Arbeit mit erm\"oglicht haben. Auf wissenschaftlicher Seite m\"ochte ich
zun\"achst Rosario Fazio f\"ur die langj\"ahrige gute Zusammenarbeit sowie
f\"ur das angenehme Arbeitsklima am DMFCI in Catania danken, 
welches zu schaffen eine nicht zu vernachl\"assigende Leistung darstellte.
Neben allen Mitgliedern der Theoriegruppe um Saro Fazio und Pino Falci,
mit denen ich viele Jahre angenehmer Zusammenarbeit mit vielen interessanten
Diskussionen und gemeinsamen Ver\"offentlichungen teilen durfte,
m\"ochte ich ganz besonders Luigi Amico danken f\"ur seine gute 
Freundschaft und die langj\"ahrige gute Zusammenarbeit mit vielen 
intensiven Diskussionen.
Weiter geht mein Dank an die Mitglieder des
Instituts f\"ur theoretische Physik hier an der Universit\"at Hannover,
insbesondere an Holger Frahm f\"ur eine angenehme und
unkomplizierte Arbeitsathmosph\"are und Zusammenarbeit, 
die mir die M\"oglichkeit gab, meine Arbeit im Bereich der 
Quanteninformation konzentriert bis zu diesem Punkte voranzutreiben.
Den Kollegen, die an den hier zugrundeliegenden Publikationen mitgewirkt
haben, gilt nochmals mein besonderer Dank f\"ur die dadurch entstandenen
und noch wachsenden spannenden Projekte, welche durch viel Arbeit, ganz besonders 
aber aus vielen produktiven Diskussionen geboren wurden.}

{\large
Zuguterletzt gilt mein ganz spezieller Dank meiner Familie,
allen voran Rita, welche so selbstverst\"andlich 
den Sprung vom Meer im S\"uden bis an die Leine mitmachte,
und die meine Forscherseele nach vollen Kr\"aften unterst\"utzt,
und ganz speziell auch meinen beiden kleinen T\"ochtern, Sophie und Carolina, 
welche am Abend angesprungen kommen, wie nach einer langen Reise;
sie bilden zusammen und ganz besonders dann, wenn mal wieder so gar nichts 
funktionieren will, den tragenden Ruhepfeiler und 
die wirksamste Tiramis{\'u}\footnote{Jede der so vielen
praktisch unwiderstehlichen Pasticcerie siciliane wurde dabei 
mit eingerechnet!} 
unterm Himmel, wenngleich sie es zum Teil noch gar nicht wissen.
}

\newpage

${}$
\newpage

\begin{center}
{\huge \bf Zusammenfassung}
\end{center}

\vspace*{3mm}

Die Anwesenheit verschr\"ankter Quantenzust\"ande f\"uhrte
zu Unbehagen und Zweifeln an der Quantentheorie, ob der Nichtlokalit\"at 
die sie in sich tragen. 
Da Lokalit\"at einer der Hauptpfeiler physikalischer Gedankengeb\"aude war und noch
ist, wurde die Annahme der Unvollst\"andigkeit der Quantentheorie und
damit die Existenz versteckter lokaler Variablen geboren.
Es war die Bahnbrechende Arbeit von John Bell, die es m\"oglich machte,
anhand von sogenannten \glqq Bellungleichungen\grqq\/ die Vorhersagen
der Quantentheorie von denen einer Theorie mit versteckten Variablen zu unterscheiden;
bislang sind derartige Messungen zugunsten der Quantentheorie ausgefallen.
Es ist seit der Gr\"undung der Quanteninformationstheorie, da\3 die Verschr\"anktheit von Zust\"anden
- das \glqq Entanglement\grqq\/ - wieder in den Mittelpunkt des Interesses ger\"uckt ist,
und zwar als Ressource f\"ur die Ausf\"uhrung klassisch unm\"oglicher Prozesse, wie 
etwa der Teleportation.
Dieser Aspekt f\"uhrte unausweichlich zu dem Bestreben wissenschaftlicher Arbeiten, diese 
Ressource klassifizieren und nat\"urlich auch quantifizieren zu wollen.
Zu diesem Zwecke sind Minimalkriterien an ein Ma\3 f\"ur Entanglement erarbeitet worden,
welche auf das Konzept des \glqq Entanglement Monotones\grqq\/ f\"uhrten.
Dieses Fundament motivierte nachfolgend lebhafte wissenschaftliche Aktivit\"at mit
Fokus auf das Entanglement von Bipartitionen, welche wichtige Kriterien wie  
den Schmidt-Rang, die von Neumann Entropie und die \glqq Concurrence\grqq\/
hervorbrachten. Der Erfolg bei bipartiten Systemen rief nach Erweiterungen der dort
gefundenen Resultate auf multipartite Systeme; jedoch erwies sich dieses neue Feld 
als weitaus komplizierter. Ausschlaggebend hierf\"ur ist letztlich die parallele
Existenz verschiedener Entangelmentklassen bez\"uglich stochastisch lokaler Transformationen
begleitet von klassischer Kommunikation (SLOCC).

Von vor etwa zehn Jahren entstammte die Idee, da\3 die Quanteninformationstheorie
das Potential haben k\"onnte, ein tieferes Verst\"andnis von komplexen Ph\"anomenen
im Bereich der kondensierten Materie oder der Quantenfeldtheorie zu erlangen.
Tats\"achlich f\"uhrte die darauffolgende intensive Forschungsarbeit unter dieser Pr\"amisse
auf beiden Gebieten zu einer wechselseitigen Befruchtung.
Von der dramatisch anwachsenden Intensit\"at wissenschaftlicher Arbeit auf dem \"Uberlapp
beider Gebiete profitierten beide Seiten. Als besonders relevant f\"ur die vorliegende Arbeit sei hier
die Untersuchung von Entanglementaspekten in der N\"ahe quantenkritischer Punkte zu nennen;
jedoch f\"uhrte die Sichtweise der Quanteninformationstheorie auch schon zu
wichtigen Modifikationen numerischer Simulationsmethoden im Bereiche der kondensierten Materie,
wie beispielsweise der DMRG.
Desweiteren ist der vielerseits ertr\"aumte Quantenrechner letztendlich ein gro\3es System
von Quanteninformationseinheiten (z.B. Qubits), f\"ur welche lokale Operationen, aber auch
paarweise Wechselwirkungen untereinander auf kontrollierte Weise manipuliert werden k\"onnen
m\"ussen. Daher sind z.B. Spinketten als Quantenregister, also als Tr\"ager von Quanteninformation,
bzw. als Quantenkanal vorgeschlagen worden. Im letzteren Falle w\"urde die nat\"urlich
gegebene hamilton'sche Zeitentwicklung zum Transport von Quantenbits ausgenutzt werden wollen.

Viele Arbeiten untersuchten also die Dynamik von Entanglement in Systemen kondensierter
Materie z.B. unter dem Aspekt optimaler Daten\"ubertragung, oder aber des Entanglementgehalts
von Grundzust\"anden quantenkritischer Modelle. 
Ein wichtiges und praktisch einhelliges Resultat der Untersuchungen der letzteren Kategorie ist,
da\3 das f\"ur Quantenphasen\"uberg\"ange wichtige Entanglement vornehmlich multipartiter Natur ist.
Diese Erkenntnis entfesselt eine Plethora ungel\"oster Probleme, welche bis in Bereiche der
Invariantentheorie reichen.
Ohne klare Vorstellung, welche Entanglementklasse f\"ur bestimmte komplexe Ph\"anomene
von Wichtigkeit sein k\"onnte, wird ein m\"oglicher Zusammenhang nur schwer hergestellt werden 
k\"onnen; aber daf\"ur w\"are eine bekannte Klassifizierung des Entanglements Voraussetzung.
Dieses Manko f\"uhrt  dazu, da\3 zun\"achst leicht berechenbare Gr\"o\3en analysiert werden,
welche aber dennoch gewisse Schl\"usse \"uber das Entanglement im betrachteten System 
zulassen. Diese erzwungen pragmatische Herangehensweise ist zwar wichtig;
sie l\"a\3t jedoch viele Facetten des Problems aus, und d\"urfte daher auf lange Sicht unzureichend sein.

Die vorliegende Arbeit ist eine Zusammenfassung der in der vorhergehenden Erkl\"arung
enthaltenen Liste von Publikationen. Nach einer Einf\"uhrung in die
meistgenutzten Entanglementma\3e, greift sie an das Problem des mutma\3lichen 
Zusammenhangs zwischen Entanglement und Quantenphasen\"uberg\"angen an und zitiert
eine Reihe von Arbeiten zu diesem Thema als Beleg f\"ur die Relevanz multipartiten 
Entanglements. Darauffolgend wird das Problem der Quantifizierung und Klassifizierung
\glqq genuin multipartiten Entanglements\grqq\/ formuliert und angegangen.
Die Schl\"usselerkenntnis hierzu ist, die $SL(2,\CC)$ samt Qubitpermutationen als
Invarianzgruppe zu identifizieren. Lokale antilineare Operatoren mit verschwindenden Erwartungswerten
auf dem gesamten lokalen Hilbertraum werden als Bausteine f\"ur solche Ma\3e vorgestellt.
Auf diese Weise konnte eine vollst\"andige Klassifizierung vierpartiten Entanglements
erfolgen, aber auch Ma\3e f\"ur echt multipartites Entanglement f\"ur eine beliebige Anzahl
von Qubits sind in Reichweite.
Die konstruierten Ma\3e sind zun\"achst nur wohldefiniert auf reinen Zust\"anden;
die Erweiterung auf gemischte Zust\"ande mittels des sogenannten \glqq convex roof\grqq\/
stellt eines der ungel\"osten Probleme dar. 
Auf dem Wege zu dessen L\"osung konnten Gemische zweier bestimmter tripartiter Zust\"ande
analytisch behandelt werden; einige der Ergebnisse dieser Arbeit lassen sich sogar auf 
beliebige Rang-zwei Gemische und f\"ur beliebige Anzahl der Qubits \"ubertragen.

Die Konstruktion mit lokalen invarianten Operatoren 
ist vom physikalischen Standpunkt besonders sinnvoll, da
sie den Grundstein daf\"ur legt, die Entanglementma\3e durch Korrelationsfunktionen
auszudr\"ucken. Dazu konnte eine eins-zu-eins-Beziehung von Erwartungswerten antilinearer 
hermitescher Operatoren mit Erwartungswerten eindeutig zugeordneter 
linearer hermitescher Operatoren hergestellt werden. 
Absch\"atzungen f\"ur das convex roof schwach gemischter Zust\"ande
werden dann eine direkte Ankn\"upfung an das Experiment liefern.
\newpage

\tableofcontents

${}$
\newpage
${}$
\newpage

\chapter{Introduction}
Entanglement has been always considered as the counter intuitive part and
``spooky'' non-locality inherent to quantum mechanics~\cite{Bell87}. 
It apparently contradicts
one of the basic pillars of physics - locality - 
and gave rise to severe skepticism for several decades. 
It was after the seminal contribution of John Bell that the fundamental 
questions related to the existence of entangled states could 
be tested experimentally. Under fairly general assumptions, Bell derived a set 
of inequalities for correlated measurements of two physical observables that any
local theory should obey.  The overwhelming majority of experiments done so far 
are in favor of quantum mechanics thus demonstrating that quantum entanglement 
is physical reality~\cite{Peres93}\footnote{There are states that do not violate Bell 
inequalities and nevertheless are entangled~\cite{Methot06}.}.

Entanglement has gained renewed interest with the 
development of quantum information science~\cite{nielsen}. 
In its framework, quantum entanglement is viewed at as
a precious resource for quantum information processing. It is e.g. believed 
to be the main ingredient to the quantum speed-up in quantum computation and 
communication. Moreover several quantum protocols, as teleportation~\cite{Bennett93}, 
can be realized exclusively with the help of entangled states.

The role of entanglement as a resource in quantum information has stimulated 
intensive research that tries to unveil both its qualitative and quantitative 
aspects~\cite{Bruss01,bengtsson,Th-Eisert,Wootters01,Plenio98,Plenio07}. 
Many criteria have been proposed to 
distinguish whether a pure state is entangled or not, as for example 
the Schmidt rank and the von Neumann entropy, and
necessary requirements to be satisfied by an  entanglement measure 
have been elaborated and have lead to the notion of an entanglement 
monotone~\cite{MONOTONES}. 
Since then, a substantial bulk of work appeared on entanglement monotones for bipartite systems, 
in particular for the case of qubits. The success in the bipartite case for qubits asked 
for extensions to the multipartite case, but 
the situation proved to be far more complicated: different classes 
of entanglement 
occur, which are inequivalent not only under deterministic local 
operations and classical communication (LOCC), but even under their 
stochastic analogue (SLOCC)~\cite{SLOCC}. 

During the last decade it has been suggested
that quantum information science might bear the potential to give further insight
into areas of physics as condensed matter or quantum 
field theory~\cite{preskill00}.
Indeed has the growing interest of the quantum information community 
in systems from condensed matter stimulated an 
exciting cross-fertilization between the two areas;
the amount of work at the interface between condensed matter physics
and quantum information theory has grown tremendously during the 
last few years, shining light on many different aspects of both subjects.
Methods developed in quantum information have proved to be
useful also in the analysis of many-body systems. 
At the same time, the experience 
built up over the years in condensed matter physics is valuable for finding new 
protocols for 
quantum computation and communication: at the end, a quantum computer will be  a 
many-body system where, differently from the `traditional' situation, 
the Hamiltonian must be manipulated in a controlled manner. 
Spin networks have been proposed as quantum channels~\cite{bose03} where 
the collective dynamics of their low lying excitations is exploited for 
transporting quantum information. 
But tools from quantum information theory also start influencing
numerical methods as the density matrix renormalization group and the 
design of new efficient simulation strategies for many-body systems 
(see for example~\cite{dmrg1,dmrg2,dmrg3}). 
Of particular interest for this thesis will be the extensive analysis of entanglement 
in quantum critical models~\cite{OstNat,Osborne02,GVidal03}. 

One important conclusion from the enormous bulk of work concerned with
entanglement at quantum phase transitions is that multipartite
quantum correlations are typically playing a dominant role~\cite{AFO07}.
This establishes an interconnection with the field of invariant theory,
where the quantification and classification of multipartite entanglement
provides with a plethora of open problems also interesting in mathematics.
Without a clear perspective of which multipartite quantum correlations
might have relevance for certain complex phenomena in condensed matter
physics and - most importantly - in absence of a full classification of entanglement,
essentially those measures are investigated that can be easily computed.
Though important in its own right, this pragmatic approach waives 
the main scope behind such an analysis,
namely the understanding of the underlying entanglement pattern
and its interconnection with complex physical phenomena.   

Most of this thesis represents a summary of a selection of  work
published during the last few years~\cite{OstNat,amico04,Amico-Osterloh,OsterlohSPIE,OS04,OS05,oster06,LOSU,AFO07}.
It will start with an overview over largely employed entanglement
measures followed by a selection of results of their analysis
for condensed matter systems emphasizing the relevance of multipartite
entanglement. 
Thereafter, the concept of genuine multipartite entanglement is introduced
and an approach for the construction of 
genuine multipartite entanglement measures is presented.
Relevant new features appearing in the multipartite case, 
as compared to bipartite measures, are highlighted, together with an outline
of how to measure multipartite entanglement in pure states in the laboratory.

\chapter{Pairwise and Bipartite Entanglement}
The problem of measuring entanglement is a vast and lively field of 
research in its own. Numerous different methods have been proposed for its quantification
In this Section we do not attempt to give an exhaustive
review of the field.
Rather do we want to introduce those measures that are largely being used to 
quantify entanglement in many-body systems. Comprehensive overviews of 
entanglement measures can be found in~\cite{Bruss01,bengtsson,Th-Eisert,
Wootters01,Plenio98,Plenio07,Horodecki07}. 
Also a method for detecting entanglement is outlined that is based on
entanglement witnesses.

\section{Bipartite entanglement in pure states}
\label{measure-pairwise-pure}

Bipartite entanglement of pure states is conceptually well understood, 
although quantifying it for local dimensions higher than two still bears 
theoretical challenges~\cite{ORDERING,Horodecki07}. 
A pure bipartite state is not entangled if and only 
if it can be written as a tensor product of pure states of the parts.
It is an important fact with this respect that for every pure bipartite 
state $\ket{\psi_{AB}}$ (with the two parts, $A$ and $B$),
two orthonormal bases $\{\ket{\psi_{A,i}}\}$ and $\{\ket{\phi_{B,j}}\}$ exist
such that $\ket{\psi_{AB}}$ can be written as
\beq
\ket{\psi_{AB}}=\sum_{i}\alpha_{i}\ket{\psi_{A,i}}\ket{\phi_{B,i}}
\label{Schmidt}
\eeq
where $\alpha_i$ are positive coefficients.
This decomposition is called the Schmidt decomposition
and the particular basis coincide with the eigenbasis of the corresponding
reduced density operators
\nbeqa
\rho_B = \trace_{A}(\ket{\psi_{AB}})&=&\sum_{i} \alpha_{i}^2\ket{\psi_{B,i}}
\bra{\psi_{B,i}}  \; , \\
\rho_A = \trace_{B}(\ket{\psi_{AB}})&=&\sum_{i} \alpha_{i}^2\ket{\phi_{A,i}}
\bra{\phi_{A,i}} \; . 
\neeqa
The density operators $\rho_A$ and $\rho_B$ have common 
spectrum, in particular are they equally mixed. 
Since only product states lead to pure reduced density matrices,
a measure for their mixedness points a way towards quantifying entanglement in this 
case. Given the state $\ket{\psi_{AB}}$, we can thus take its Schmidt 
decomposition, Eq.(\ref{Schmidt}), and use a suitable function of the 
$\alpha_i$ to quantify the entanglement. 

It is interesting that an entanglement measure $E$ is fixed uniquely 
 after imposing the following conditions
\begin{enumerate}
\item  $E$ is invariant under local unitary operations ($\Rightarrow$ 
       $E$ is indeed a function of the $\alpha_i$'s only).

\item  $E$ is continuous (in a certain sense also in the asymptotic limit of infinite copies 
       of the state; see e.g. Ref.~\cite{Plenio07}).

\item  $E$ is additive, when several copies of the system are present:\\
       $E(\ket{\psi_{AB}}\otimes \ket{\phi_{AB}}) = 
       E(\ket{\psi_{AB}})+E(\ket{\phi_{AB}})$.
\end{enumerate}
The unique measure of entanglement satisfying 
all the above conditions is the von Neumann entropy of 
the reduced density matrices
\begin{equation}
S(\rho_A)= S(\rho_B)= -\sum_i \alpha_i^2 \log(\alpha_i^2) \; ,
\end{equation}
this is just the Shannon entropy of the moduli squared of the
Schmidt coefficients. In other words: one possible answer on the question
of how entangled a bipartite pure state is, 
can be given by the von Neumann entropy of (either of)
the reduced density matrices. 
The amount of entanglement is generally difficult to define once we are 
away from bipartite states, but in several cases we can still gain some 
insight into many-party entanglement if one considers different 
bipartitions of a multipartite system.  In particular 
if no reduced density matrix is pure, 
then the state is called globally entangled.
 
It is worth to notice that 
a variety of purity measures are admissible when 
the third condition on additivity is omitted. In principle, there are infinitely many
measures for the mixedness of a density matrix; two of them will typically
lead to a different ordering when the Hilbert space of the parts has a 
dimension larger than two. This is essentially equivalent
to saying that different inequivalent classes of entanglement exist in these
cases.   
In contrast, if we trace out one of two {\bf qubits} 
the corresponding reduced density matrix $\rho_A$ contains only a single 
independent and unitarily invariant parameter: its smallest eigenvalue. 
This implies that 
each monotonic function $[0,1/2]\mapsto [0,1]$ of this eigenvalue
can be used as an entanglement measure. 
Though, also here an infinity of different mixedness measures
exists, here all lead to the same ordering of states with respect to their 
entanglement, and in this sense all are equivalent.
A relevant example is the (one-) tangle~\cite{Coffman00}
\begin{equation}
\tau_1[\rho_A]:=4 {\rm det} \rho_A\; .
\label{onetangle}
\end{equation}
By expressing $\rho_A$ in terms of spin form factors
\beq
\rho_A=
\Matrix{cc}{\frac{1}{2}+\expect{S^z}&\expect{S^x}-i\expect{S^y}\\
\expect{S^x}+i\expect{S^y}&\frac{1}{2}-\expect{S^z}}\; ,
\label{rho1} 
\eeq
where $\expect{S^{\alpha}} = tr_B(\rho_A S^{\alpha})$ and 
$S^{\alpha}=\frac{1}{2} \sigma^\alpha$, 
$\sigma^\alpha$ $\{\alpha=x,y,z\}$ being the Pauli matrices, it follows that 
$$
\tau_1[\rho_A]=1-
4(\expect{S^x}^2+\expect{S^y}^2+\expect{S^z}^2)\; .
$$
For a pure state $\ket{\psi_{AB}}$ of two qubits the relation 
\beq
\tau_1\equiv
|\bra{\psi^*}\sigma^y_{A} \otimes \sigma^y_{B}\ket{\psi}|^2=:C[\ket{\psi_{AB}}]^2=:\tau_2
\eeq
applies, where $C$ is called the pure state concurrence~\cite{Hill97,Wootters98},
and $*$ indicates the complex conjugation in the eigenbasis of $\sigma^z$.
It is worth emphasizing already here that all measures of pairwise qubit
entanglement can hence be expressed in terms of the modulus squared
of the expectation value of an antilinear operator.
This innocent looking detail enhances the minimally required
invariance with respect to local $SU(2)$ transformations to the 
local invariance group $SL(2)$. The latter is known to be the relevant
invariance group for the classification of entanglement into
SLOCC classes, where generalized local measurements are admitted in
a probabilistic way~\cite{Duer00}.
This observation will be a key element, paving the way
towards the quantification and classification of multipartite entanglement.

The von Neumann entropy can be expressed as a function of the (one-) tangle
$$\displaystyle{
S[\rho_A]=h\left(\frac{1}{2}\left(1+\sqrt{1-\tau_1[\rho_A]}\right) \right)}
$$
where
$h(x)=: -x \log_2 x - (1-x) \log_2 (1-x)$ is the binary entropy.
Both the tangle and the concurrence lead to the important {\em monogamy} 
inequality~\cite{Coffman00,Osborne06} which will be discussed in the 
next Section.

\section{Pairwise qubit entanglement in mixed states}
\label{measure-pairwise-mixed}

Subsystems of a many-body (pure) state will generally be in a mixed state,
and then even different concepts of entanglement do exist.
Three important representatives are the entanglement cost $E_C$,
the distillable entanglement $E_D$ (both defined in Ref.~\cite{Bennett96}) and the 
entanglement of formation $E_F$~\cite{BennettDiVincenzo96}.
Whereas $E_D$ and $E_C$ are asymptotic limits of multi-copy extraction
probabilities of Bell states and creation from such states, the entanglement of
formation is the amount of pure state entanglement needed to create
a single copy of the mixed state. 
Very recently, a proof of the full additivity of $E_F$ has been 
presented~\cite{AdditiveEoF}, which implies that for bipartite systems
both concepts coincide (see e.g.~\cite{VidalECost02}), i.e. $E_D=E_C$. 
The conceptual difficulty behind the calculation of $E_F$ lies
in the infinite number of possible decompositions of a density matrix.
Therefore, even knowing how to quantify bipartite 
entanglement in pure states, we cannot simply apply this knowledge 
to mixed states in terms of an average over the mixtures of pure state entanglement. 
The problem is that two decompositions of the same density matrix 
usually lead to a different average entanglement. 
Which one do we choose? 
It turns out that we must take the minimum over all possible decompositions, 
simply because if there is a decomposition where the average is zero, 
then this state can be created locally without need of any entangled pure
state, and therefore $E_F=0$. 
The same conclusion can be drawn from the requirement that entanglement must not
increase on average by means of local operations including classical communication 
(LOCC). A minimal set of requirements every entanglement measure has to fulfill has lead to the 
notion of an {\em entanglement monotone}~\cite{MONOTONES}.

The entanglement of formation of a state ${\rho}$ is 
therefore defined as 
\begin{eqnarray}
E_{F}(\rho):= \min \sum_j p_j S(\rho_{A,j}) \; , \label{ef}
\end{eqnarray}
where the minimum is taken over all realizations of the state 
$\rho_{AB} = \sum_j p_j |{\psi_j}\rangle\langle{\psi_j}|$, and $S(\rho_{A,j})$ 
is the von Neumann entropy of the reduced density matrix 
$\rho_{A,j}:=\trace_B \ket{\psi_j}\bra{\psi_j}$.
Eq.(\ref{ef}) is the so-called {\em convex roof} of the entanglement of 
formation for pure states, and a decomposition leading to this
convex roof value is called an {\em optimal decomposition}.

For systems of two qubits, an analytic expression for $E_F$ does exist
and it is given by
\begin{equation}
\label{EoF}
E_F (\rho) = -\sum_{\sigma = \pm}\frac{\sqrt{1+\sigma 
C^2(\rho)}}{2}\ln\frac{\sqrt{1+\sigma C^2(\rho)}}{2}
\end{equation}
where $C(\rho)$ is the convex roof of the pure state concurrence~\cite{Wootters98,Wootters01} 
which has been defined in the previous section. 
Its convex roof extension is encoded in the positive Hermitean matrix 
\begin{equation}
  \label{eq:productmatrix}
  R\equiv\sqrt{\rho}\tilde{\rho}\sqrt{\rho} =\sqrt{\rho}(\sigma^y\otimes\sigma^y)
\rho^*(\sigma^y\otimes\sigma^y)\sqrt{\rho}\; ,
\end{equation}
with eigenvalues $\lambda_1^2\geq \dots \geq \lambda_4^2$ 
in the following way\footnote{$\tilde{\rho} := 
(\sigma^y\otimes \sigma^y)
\rho^*(\sigma^y\otimes \sigma^y)$ is the tilde-conjugate 
of the density matrix, which here coincides with its time reversal.}
\begin{equation}
  \label{eq:concurrence}
  C = \mbox{max}\{\lambda_1-\lambda_2-\lambda_3-\lambda_4,0\}\; .
\end{equation}
As the entanglement of formation is a monotonous function of the concurrence, 
also $C$ itself or its square $\tau_2$ can be used as entanglement measures. This
is possible due to a curious peculiarity of two-qubit systems: namely 
that a continuous variety of optimal decompositions exist~\cite{Wootters98}.
The concurrence $C$ and the tangle $\tau_1$ both range 
from $0$ (no entanglement) to $1$.

By virtue of (\ref{eq:productmatrix}) and (\ref{eq:concurrence}), 
the concurrence in a spin-1/2 chain can be computed
in terms of up to two-point spin correlation functions.  
For simplicity we consider a case where the model has a 
parity symmetry~\footnote{In this case all the components of the wave 
function have an even (or odd) number of flipped spins}. 
For this case the reduced density matrix $\rho_{ij}$ for spins placed at 
sites $i$ and $j$ assumes a simple form in the basis 
$\{\ket{00},\ket{01},\ket{10},\ket{11}\}$
\footnote{when discussing qubits or spin-1/2 systems
both notations $\ket{0},\ket{1}$ and $\ket{\uparrow},\ket{\downarrow}$ 
will be used for the eigenstates of $S^z$.} 
\begin{equation}
\label{rho:non-equilibrium}
\rho_{ij}^{(2)}\ =\ \Matrix{cccc}{
a_{ij}&0&0&c_{ij}\\
0&x_{ij}&z_{ij}&0\\
0&z^*_{ij}&y_{ij}&0\\
c^*_{ij}&0&0&b_{ij}\\
}\, ,
\end{equation}
with real $a, b, x, y$, and complex $c, z$. The concurrence results to be 
\beq
C_{ij}=2\max\{0,|c_{ij}|-\sqrt{x_{ij}y_{ij}},|z_{ij}|-\sqrt{a_{ij}b_{ij}}\}\; .
\eeq 
For  translational invariant systems: $x=y$; for real Hamiltonians
and stationary states: $c,z \in \RR$. Each entry of the matrix $\rho_{ij}$ is 
then simply related to one- and two-point correlation functions,
\begin{equation}
\label{C-of-corrs}
C_{ij}=2\max\left\{0,C_{ij}^I,C_{ij}^{II}\right\}\;.
\end{equation}
where \begin{eqnarray}
C^I_{ij}&=&|g_{ij}^{xx}+g_{ij}^{yy}|-\sqrt{\left(\frac{1}{4}+g^{zz}_{ij}\right)^2-M_z^2}~,
\label{e.C'r}\\
C^{II}_{ij}&=&|g_{ij}^{xx}-g_{ij}^{yy}|+g_{ij}^{zz}-\frac{1}{4}~,
\label{e.C''r}
\end{eqnarray} 
with $g_{ij}^{\alpha\alpha}=\langle S^\alpha_i S^\alpha_{j} \rangle$
and $M_z=\langle S^z \rangle$ (assuming translational invariance).
A state with dominant fidelity of parallel and anti-parallel Bell states is 
characterized by dominant $C^I$ and $C^{II}$, respectively. 
This was shown in~\cite{Fubini06}, 
where the concurrence was expressed in terms of the {\em fully entangled 
fraction} as defined in~\cite{BennettDiVincenzo96}.

The importance of the tangle and the concurrence is due to the {\em monogamy} 
inequality derived in~\cite{Coffman00} for three qubits. 
This inequality has been 
proved to hold also for n-qubits system~\cite{Osborne06}. In the case of 
many-qubits (the tangle may depend on the site $i$)
it reads 
\begin{equation}
        \sum_{j\neq i} C^2_{ij} \leq  \tau_{1,i} \; .
\end{equation}
The so called {\em residual tangle} $\tau_{1,i}-\sum_{j\neq i} C^2_{ij}$, is 
then a measure for multipartite entanglement~\cite{Coffman00,Osborne06}
not stored in pairs of qubits only.
We finally mention that the antilinear form of the concurrence was the key
for the first explicit construction of a convex roof, and hence its 
extension to mixed states~\cite{Hill97,Wootters98,Uhlmann00}. 

Another measure of entanglement we mention is the relative entropy 
of entanglement~\cite{RelativeEntropy}. It can be applied to any number of 
qubits in principle (or any dimension of the local Hilbert space).
It is formally defined as
$
E({\sigma}):= \min_{\rho \in {\cal D}}\,\,\, S(\sigma || \rho)
$,
where 
$S(\sigma || \rho) = \trace \sigma \left[\ln \sigma - \ln \rho \right]$
is the quantum relative entropy. This relative entropy of entanglement
quantifies the entanglement in $\sigma$ by its distance from the set ${\cal D}$
of separable states. 
The main difficulty in computing this measure is to find the 
disentangled state closest to $\rho$. This is in general a difficult task,
even for two qubits. In the presence of certain symmetries 
- which is the case for e.g. eigenstates of certain models - an analytical 
access is possible. In these cases, the relative entropy of entanglement  
becomes a very useful tool. The relative entropy reduces to the 
entanglement entropy in the case of pure bi-partite states; this also 
means that its convex roof extension coincides with the 
entanglement of formation, and is readily deduced from the 
concurrence~\cite{Wootters98}. 

\section{Localizable entanglement}
\label{locsection}

A different approach to entanglement in many-body systems arises from the 
quest to swap or transmute different types of multipartite entanglement
into pairwise entanglement between two parties by means of 
generalized measures on the rest of the system. In a system 
of interacting spins on a lattice one could then try to maximize the 
entanglement between two spins (at positions $i$ and $j$) by performing 
measurements on all the others. The system is then partitioned in three 
regions: the sites $i$, $j$  and the rest of the lattice.
This concentrated pairwise entanglement can then
be used  e.g. for quantum information processing.
A standard example is that the three qubit GHZ state 
$(1/\sqrt{2})(\ket{000}+\ket{111})$ 
after a projective measure in $x$-direction on one of the sites 
is transformed into a two qubit Bell state.

The concept of localizable entanglement has been introduced 
in~\cite{verstraete04,popp05}. 
It is defined as the maximal amount of entanglement that can be 
localized, on average, by doing local measurements in the rest of the 
system\footnote{These operations in principle need not to be local 
operations in terms of the multipartite setting of single spins on all 
the chain.}.  
In the case of $N$ parties, the possible outcomes of the measurements 
on the remaining $N-2$ particles are pure states $|\psi_s \rangle$ 
with corresponding probabilities $p_s$. The localizable entanglement 
$E_{loc}$ on the sites $i$ and $j$ is defined as the maximum of the average entanglement over 
all possible outcome states $\ket{\psi_s}_{ij}$
\begin{equation}
E_{loc}(i,j) = \mbox{sup}_{\cal{E}} \sum_s p_s E(\ket{\psi_s}_{ij})
\end{equation}
where ${\cal E}$ is the set of all possible outcomes $(p_s, |\psi_s \rangle)$ of the measurements,
and $E$ is 
the chosen measure of entanglement of a pure state of two qubits 
(e.g. the concurrence). 
Although very difficult to compute, lower and upper bounds have 
been found which allow to deduce a number of consequences for this quantity.  

An upper bound to the localizable entanglement is given by the 
entanglement of assistance~\cite{laustsen03} obtained from 
localizable entanglement when also global and joint measurements 
were allowed on the $N-2$ spins~\footnote{The entanglement of assistance is 
defined as the maximum average entanglement among the pure 
state realizations of the state under consideration~\cite{Bennett96}}.
A lower bound of the localizable entanglement comes from the 
following theorem~\cite{verstraete04} 
\begin{theorem}
Given a (pure or mixed) state of N qubits with reduced 
correlations $Q_{ij}^{\alpha,\beta} = 
\langle S_{i}^{\alpha}S_{j}^{\beta}\rangle - 
\langle S_{i}^{\alpha}\rangle \langle S_{j}^{\beta}\rangle$ 
between the  spins $i$ and $j$ and 
directions $\alpha$ and $\beta$ then there always exist directions in 
which one can measure the other spins such that this correlation do not 
decrease, on average.
\end{theorem}
It then follows that a lower bound to localizable entanglement is fixed 
by the maximal correlation function between the two parties 
(one of the various spin-spin correlation functions 
$Q_{ij}^{\alpha,\beta}$)\footnote{It has been argued recently~\cite{Gour05,Gour06} 
that in order to extend the entanglement of assistance and 
the localizable entanglement to being an entanglement monotone~\cite{MONOTONES} one should
admit also local operations including classical communication on the 
extracted two spins, this was named entanglement of collaboration.}.

\section{Entanglement witnesses} 

It is important to realize that not just the quantification of 
many-party entanglement is a difficult task; it is an open 
problem to tell in general, whether a state of $n$ parties is separable or not
although a formal solution of the problem can be written in several forms. 
It is therefore of great value to have a tool that is able to merely
certify if a certain state is entangled. An entanglement witness $W$ 
is a Hermitean operator which is able to detect entanglement in a state. 
The basic idea is
that the expectation value of the witness $W$ for the state $\rho$
under consideration exceeds certain bounds only when $\rho$ is entangled. 
An expectation value of $W$ within this bound however
does not guarantee that the state is separable. Nonetheless, this is a very 
appealing method also from an experimental point of view, since it is 
sometimes possible to relate the presence of the entanglement to the 
measurement of few observables.

Simple geometric ideas help to explain the  witness 
operator $W$ at work. Let $\mathcal{T}$ be the set of all density matrices 
and let $\mathcal{E}$ and $\mathcal{D}$ be the subsets of entangled and 
separable states, respectively. 
The convexity of $\mathcal{D}$ is a key property for witnessing 
entanglement\footnote{This is based on the Hahn-Banach separation 
theorem, stating that given a convex set and a point outside there exists a 
plane that separates the point from the set.}.
The entanglement witness is then an operator 
defining a hyper-plane which separates a given entangled state from the set of 
separable states. The main scope of this geometric approach is then 
to optimize the witness operator~\cite{LewensteinOptWit} or to replace 
the hyper-plane by a curved manifold, tangent to the set of separable 
states~\cite{witguehne}~\footnote{For other geometric aspects of entanglement 
see~\cite{Klyachko,bengtsson,Leinaas06}.}.
We have the freedom to choose $W$ such that 
$$
\trace(\rho_D W)\leq 0
$$ 
for all disentangled states $\rho_D\in{\cal D}$.
Then,  
$$
\trace(\rho W)>0
$$ 
implies that $\rho$ is entangled.
A caveat is that the concept of a witness is not invariant 
under local unitary operations (see e.g.~\cite{FoolingWits}).

Entanglement witnesses are a special case of a more general concept, 
namely that of positive maps. These are injective superoperators
on the subset of positive operators.
When we now think of superoperators that act non-trivially only on a sub-Hilbert space,
then we may ask the question whether a positive map on the subspace is 
also positive when acting on the whole space. Maps that remain positive 
also on the extended space are called {\em completely positive maps}.
The Hermitean time evolution of a density matrix is an example for a 
completely positive map. Positive but {\bf not} completely positive 
maps are important for entanglement theory~\cite{Horodecki96,Jamiolkowski}
\begin{theorem} 
A state $\rho_{AB}$ is entangled if and only if
a positive map $\Lambda$ exists (not completely positive) such that 
$$
(\id_A \otimes \Lambda_B) \rho_{AB} < 0\; .
$$ 
\end{theorem}
For a two dimensional local Hilbert space the situation simplifies
considerably in that any positive map $P$ can be written as
$
P=CP_1 + CP_2 T_B \; ,
$
where $CP_1$ and $CP_2$ are completely positive maps and $T_B$ is a
transposition operation on subspace $B$. This decomposition tells that for a 
system of two qubits the lack of complete positivity in a positive
map is due to a partial transposition. This partial transposition 
clearly leads to a positive operator if the state is a tensor 
product of the parts. In fact, also the opposite is true: 
a state of two qubits $\rho_{AB}$ is separable if and only if 
$\rho_{AB}^{T_B} \geq 0$ that is, its partial transposition is positive. 
This is very simple to test and it is known as the Peres-Horodecki 
criterion~\cite{Peres96,Horodecki96}. The properties of entangled 
states under partial transposition lead to a measure of entanglement
known as the {\em negativity}. The negativity $N_{AB}$  
of a bipartite state is defined as the absolute value of the sum of 
the negative eigenvalues of $\rho_{AB}^{T_A}$. The 
{\em logarithmic negativity} is then defined as
\begin{equation}
     E_N  = \log_22 (2N_{AB}  + 1).
\end{equation}
For bipartite states of two qubits, $\rho_{AB}^{T_A}$ has at most one 
negative eigenvalue~\cite{Sanpera98}. For general multipartite systems and 
higher local dimension there are entangled states with a
positive partial transpose, known as bound entangled 
states~\cite{acinbe,Horodeckibe}.

\section{Indistinguishable particles}
\label{indistparts}

There is an ongoing debate on which definition of entanglement for 
indistinguishable particles will be the most useful from a physical 
point of view. This uncertainty is responsible for the vast variety 
of quantities studied, when the entanglement of itinerant fermion and 
boson systems is discussed. 

The problem is that for indistinguishable 
particles the wave function is \mbox{(anti-)~symmetrized} and therefore the 
definition of entangled states as given in the previous Section does not 
apply. In particular, it does not make sense to consider each individual 
particle as parts of the partition of the system.
Following~\cite{Ghirardi02,Ghirardi03} one can address the 
problem of defining entanglement in an ensemble of indistinguishable 
particles by seeing if one can attribute to each of the subsystems
a complete set of measurable properties, e.g. momenta for free pointless particles.
Quantum states satisfying the above requirement are precisely 
the (anti-) symmetrization of a product state of (fermions) bosons, 
and represent the separable states for indistinguishable particles.

There is another crucial difference 
between the entanglement of (indistinguishable) spin-1/2 particles and that of 
qubits. Let us consider two fermions on two sites. Whereas the Hilbert 
space ${\cal H}_s$ of a two-site spin lattice has dimension
${\rm dim}\, {\cal H}_s= 4$, the Hilbert space ${\cal H}_f$ for two 
fermions on the same lattice has dimension ${\rm dim}\, {\cal H}_f=6$. 
This is due to the possibility that both fermions, with opposite spins, 
can be located at the  same lattice site.
When choosing the following numbering of the states
\beqb{ccc}{numbering}
\ket{1}&=:&f^\dagger_1\ket{0}=:c^\dagger_{L,\up}\ket{0}\\
\ket{2}&=:&f^\dagger_2\ket{0}=:c^\dagger_{L,\down}\ket{0}\\
\ket{3}&=:&f^\dagger_3\ket{0}=:c^\dagger_{R,\up}\ket{0}\\
\ket{4}&=:&f^\dagger_4\ket{0}=:c^\dagger_{R,\down}\ket{0}
\eeqb 
and the definition 
$\ket{i,j} = f^\dagger_i f^\dagger_j\ket{0}$, there are Bell states 
analogous to those occurring for distinguishable
particles $(\ket{1,3}\pm\ket{2,4})/\sqrt{2}$ and 
$(\ket{1,4}\pm\ket{2,3})/\sqrt{2}$. There are however new 
entangled states, as $(\ket{1,2}\pm\ket{3,4})/\sqrt{2}$,
where both fermions take the same position.
The local Hilbert space is made of four states 
labelled by the occupation number and the spin, if singly occupied. 
The {\em site-entanglement} of indistinguishable particles is then defined
as the entanglement of the corresponding Fock states. 
It can be measured e.g. by the local von Neumann entropy. 
This quantity is the analogue to the one-tangle for 
qubits, but the local Hilbert space dimension is $4$ due to the possibility of
having empty and doubly occupied sites. 
Also the quantum mutual information~\cite{Groisman05} can be defined in this way,
quantifying the total amount (classical and quantum) of
correlations stored in a given state of a second quantized system. 

For spinless fermions a one-to-one mapping to spin-1/2 chains exists  in one spatial dimension
- the Jordan-Wigner transformation.
Due to the non-locality of the Jordan Wigner transformation,
quantitative deviations  between the entanglement of spinless fermions and 
the Jordan-Wigner relative may occur for distant sites;
as far as nearest-neighbor entanglement 
is considered, both concepts completely coincide.

Although it is known how the entanglement 
of indistinguishable particles can be quantified, as will be seen in the following part, 
the major part of the literature on second quantized systems considers the 
site-entanglement described above
or the entanglement of degrees of freedom, singled out from a suitable set 
of local quantum numbers (e.g. the spin of the particle at site $i$).
In both cases, entanglement measures for distinguishable particles (see 
Sections~\ref{sec-noninteracting} and \ref{hubbardsent}) can be used.

\subsection{Two Fermion entanglement}
\label{ferment}

In this paragraph we summarize the distinct features appearing in
the quantification and classification of Fermion entanglement.
Due to the antisymmetry under particle exchange, there is no 
Schmidt decomposition for Fermions. Nevertheless,
a Fermionic analogue to the Schmidt rank, which classifies entanglement
in bipartite systems of distinguishable particles does exist: the
so called {\em Slater rank}.
A generic state of two-electrons on two lattice sites can be written as
\beq\label{gen-ferm-state}
\ket{\omega}:=\sum_{i,j=1}^4\omega_{i,j}\ket{i,j}
\eeq
where $\omega$ is a $4\times 4$ matrix which can be assumed antisymmetric
and normalized as $\trace \omega^\dagger\omega=\frac{1}{2}$ (or
equivalently $\trace \omega^*\omega=-\frac{1}{2}$).
Since here the local entities whose entanglement
shall be studied, are the particles, unitary transformations act
on the $4$-dimensional single particle Hilbert space. 
Due to the indistinguishability of the particles,  the transformation must 
be the same for each of the particles.
Given a unitary transformation $U\in {\rm SU}(4)$ such that
$f^{'}_j:=U_{jk}f^{}_k$, the transformed state is given by
$\ket{\omega'}$ where $\omega':=U\omega U^T$.
The above unitary transformation preserves the antisymmetry,
and every pure state $\omega$ of two spin-1/2 particles on two sites can be transformed 
into the normal form
\beq
\omega_s=\Matrix{cccc}{0&z_1&0&0\\-z_1&0&0&0\\0&0&0&z_2\\0&0&-z_2&0}
\eeq
In fact, every two-particle state within a $D$-dimensional single
particle Hilbert space can be transformed into the normal form
\beqa\label{normal-ferm-form}
\omega_s &=& {\rm diag}\{Z_1,\dots,Z_r,Z_0\}\\
Z_j &=& \Matrix{cc}{0&z_j\\ -z_j&0}\\
(Z_0)_{ij} &=& 0\quad ;\qquad i,j\in\{1,\dots,D-2r\} 
\eeqa
where $r$ is then called the {\em Slater rank} of the pure 
Fermion state~\cite{Schliemann01,SchLoss01,Eckert02}.
Following the definition in the introduction to this Section, 
a pure Fermion state is entangled 
if and only if its Slater rank is larger than $1$.

It is important to notice that the above concept of entanglement only
depends on the dimension of the Hilbert space accessible to each of the
particles (this includes indistinguishable particles
on a single $D$-level system).

For electrons on an $L$-site lattice the ``local'' Hilbert space dimension
is $2L$, and the question, whether a pure state living in a 
$2L$-dimensional single particle
Hilbert space has full Slater rank, can be answered
by considering the Pfaffian of $\omega$~\cite{Caianello52,Muir60}
\beq\label{Pfaffian}
{\rm pf}[\omega]:=
\sum_{\pi\in {\cal S}_{2L}^<} {\rm sign}(\pi)\prod_{j=1}^L \omega_{\pi(2j-1),\pi(2j)}
\eeq
which is non-zero only if $\omega$ has full Slater rank $L$.
In the above definition ${\cal S}_{2L}^<$ denotes those elements $\pi$ of 
the symmetric group ${\cal S}_{2L}$ with 
ordered pairs, i.e. $\pi(2m-1)<\pi(2m)$ for all $m\leq L$
and $\pi(2k-1)<\pi(2m-1)$ for $k<m$.
Notice that relaxing the restriction to ${\cal S}_{2L}^<$
just leads to a combinatorial factor of $2^L L!$ by virtue of the
antisymmetry of $\omega$ and hence can we write
\beq\label{eps-pfaff}
{\rm pf}[\omega]=\frac{1}{2^L L!} \sum_{j_1,\dots,j_{2L}=1}^{2L}
      \eps^{j_1,\dots,j_{2L}}\omega_{j_1,j_2}\dots\omega_{j_{2L-1},j_{2L}}
\eeq
where $\eps^{j_1,\dots,j_{2L}}$ is the fully antisymmetric tensor
with $\eps^{1,2,\dots,2L}=1$.
There is a simple relation between the Pfaffian and
the determinant of an antisymmetric even-dimensional matrix:
${\rm pf}[\omega]^2=\det[\omega]$.

For the simplest case of two spin-1/2 Fermions on two lattice sites
the Pfaffian reads 
${\rm pf}[\omega]=\omega_{1,2}\omega_{3,4}-\omega_{1,3}\omega_{2,4}+
\omega_{1,4}\omega_{2,3}$.
Normalized in order to range in the interval $[0,1]$ this has been
called the Fermionic concurrence 
${\cal C}[\ket{\omega}]$~\cite{Schliemann01,SchLoss01,Eckert02} 
\beq\label{ferm-Conc}
{\cal C}[\ket{\omega}]=|\braket{\tilde{\omega}}{\omega}|=8|{\rm pf}[\omega]|
\eeq
where
\beq
\tilde{\omega}:=\frac{1}{2}\eps^{ijkl}\omega_{k,l}^*
\eeq
has been called the {\em dual} to $\omega$. Then,
$\ket{\tilde{\omega}}=:{\cal D}\ket{\omega}$
is the analogue to
the conjugated state in~\cite{Hill97,Wootters98,Uhlmann00}
leading to the concurrence for qubits.
It is important to notice that the Pfaffian in Eq.(\ref{Pfaffian})
is invariant under the complexification of ${\rm su}(2L)$,
since it is the expectation value of an antilinear operator,
namely the conjugation ${\cal D}$ for the state $\ket{\omega}$.
Since this invariant is a bilinear expression in the state coefficients,
its convex roof is readily obtained~\cite{Uhlmann00}
by means of the positive eigenvalues
$\lambda_i^2$ of the $6\times 6$ matrix
\beq\label{ferm-Concurrence}
R=\sqrt{\rho}{\cal D}\rho{\cal D}\sqrt{\rho}\; .
\eeq
The conjugation $\cal D$, expressed in the basis 
\mbox{$\{\ket{1,2},\ket{1,3},\ket{1,4},\ket{2,3},\ket{2,4},\ket{3,4}\}$}
(see Eq.(\ref{numbering})), takes the form
\beq\label{conjugation}
\Matrix{cccccc}{
0&0&0&0&0&1 \\
0&0&0&0&\fat{-1}&0 \\
0&0&0&\fat{1}&0&0 \\
0&0&\fat{1}&0&0&0 \\
0&\fat{-1}&0&0&0&0 \\
1&0&0&0&0&0}{\cal C}\; ,
\eeq
where ${\cal C}$ is the complex conjugation.
Notice that the center part of this matrix 
(in bold face) is precisely
$\sigma_y\otimes\sigma_y$ and indeed corresponds to
the Hilbert space of two qubits. The remaining part of the Hilbert space
gives rise to an entanglement of different values for the 
occupation number. 
This type of entanglement was considered practically useless and 
has been referred to as the 
{\em fluffy bunny}~\cite{WisemanFluffyBunny,CiracFluffyBunny} in the literature.

For a single particle Hilbert space with dimension larger than $4$
one encounters similar complications as for two distinguishable
particles on a bipartite lattice and local Hilbert space dimension
larger than $2$, i.e. for two {\em qudits}.
This is because different classes of entanglement occur, which
are characterized by different Slater rank as opposed to 
their classification by different Schmidt rank for distinguishable particles.
The Slater rank can be obtained by looking at Pfaffian minors~\cite{Muir60}:
if the Slater rank is $r$, all Pfaffian minors of dimension larger than $2r$
are identically zero.

\subsection{Multipartite Entanglement for Fermions}

For indistinguishable particles the only classification available
up to now is to check whether or not a pure state has Slater rank 
one. Eckert {\em et al.} formulated two recursive 
lemmata~\cite{Eckert02}
\begin{lemma}
\label{Lemma4.5}
A pure $M$-Fermion state has Slater rank one if and only if
$$
\sum_{j_1,\dots,j_M=1}^{2n}\omega_{j_1,\dots,j_{M-1}}a_{j_M}
f^\dagger_{j_1}\dots f^\dagger_{j_{M-1}}
$$
has Slater rank one or zero for all $\vec{a}\in\CC^{2n}$.
\end{lemma}
\begin{lemma}\label{Lemma4.6}
A pure $M$-Fermion state has Slater rank one if and only if
\begin{eqnarray}
\sum_{i_1,\dots,i_M=1 \atop j_1,\dots,j_M=1}^{2n} [ &&
\omega_{i_1,\dots,i_{M-1}}\omega_{j_1,\dots,j_{M-1}}
a^1_{j_1}\dots a^{M-2}_{j_{M-2}}  \nonumber \\
&& \eps^{i_{M-1}i_M j_{M-1}j_M\alpha_1\dots\alpha_{2n-2}} ]=0
\end{eqnarray}
for all $\vec{a}^1,\dots,\vec{a}^{M-2}\in\CC^{2n}$ and all
$0\leq\alpha_1<\dots<\alpha_{2n-2}\leq 2n$.
\end{lemma}
They can be summarized as follows: 
let an $N$-electron state be contracted
with $N-2$ arbitrary single electron states encoded in the vectors $\vec{a}^j$
as $\vec{a}^j_kf^\dagger_k\ket{0}$ ($j=1,\dots,N-2$ and sum convention)
to a two-electron state. Then the Pfaffian of the two-electron state
is zero if and only if the original state (and hence all intermediate
states in a successive contraction) has Slater rank one.
This means that all 4-dimensional Pfaffian minors of $\omega$ are zero.

Instead of the Pfaffian of $\omega$, also the single-particle reduced 
density matrix can be considered, and its von Neumann entropy as a 
measure for the quantum entanglement has been analyzed 
in~\cite{Lizeng01,Paskauskas01}. 
It is important to remind that for distinguishable particles the local reduced 
density matrix has rank one if and only if the original state were a product.
This is no longer true for indistinguishable particles.
For an $N$-particle pure state with Slater rank one
the rank of the single-particle reduced density matrix
coincides with the number of particles, $N$.
A measure of entanglement is then
obtained only after subtraction of the constant value of the von 
Neumann entropy of a disentangled state. This must be taken into
account also for the extension of the measure to mixed states.

\subsection{``Entanglement of particles''}
\label{measures-entanglement-particles}

Entanglement in the presence of superselection rules (SSR) induced by 
particle conservation has been discussed in Refs.~\cite{Bartlett03,Wiseman03,
shuchprl,shuch}. The main difference in the concept of 
{\em entanglement of particles}~\cite{Wiseman03} from the entanglement
of indistinguishable particles as described in the preceding section
(but also to that obtained from the reduced density matrix of e.g. spin
degrees of freedom of indistinguishable particles)
consists in the projection of the Hilbert space onto a subspace
of fixed particle numbers for either part of a bipartition of the system.
The bipartition is typically chosen to be space-like, as motivated from
experimentalists or detectors sitting at distinct positions.
E.g. two experimentalists, in order to detect the
entanglement between two indistinguishable particles, must have
one particle each in their laboratory. 

This difference induced by particle number superselection 
is very subtle and shows up if multiple occupancies occur 
at single sites for Fermions with some inner degrees of freedom,
as the spin. Their contribution is finite 
for finite discrete lattices and will generally scale to zero 
in the thermodynamic limit with vanishing lattice spacing. 
Therefore both concepts of spin 
entanglement of two distant particles coincide in this limit.
Significant differences are to be expected only for finite non-dilute
systems.
It must be noted that the same restrictions imposed by SSR which change considerably the
concept of entanglement quantitatively and qualitatively,
on the other hand enable otherwise impossible protocols of quantum 
information processing~\cite{shuchprl,shuch}.

Wiseman and Vaccaro project an $N$-particle state $\ket{\psi_N}$ 
onto all possible subspaces, where the two parties have a well defined 
number $(n_A,n_B=N-n_A)$ of particles in their laboratory~\cite{Wiseman03}. 
Let $\ket{\psi[n_A]}$ be the respective projection, and let $p_{n_A}$ 
be the weight 
$\braket{\psi[n_A]}{\psi[n_A]}/\braket{\psi_N}{\psi_N}$ of this projection.
Then the entanglement of particles $E_p$ is defined as
\beq
E_p[\ket{\psi_n}] = \sum_n p_n E_M[\psi[n_A]]
\eeq
where $E_M$ is some measure of entanglement for distinguishable particles.
Although this certainly represents a definition of entanglement
appealing for experimental issues, it is sensitive only to situations,
where e.g. the two initially indistinguishable particles
eventually are separated and can be examined one-by-one by Alice and Bob.
Consequently, ``local operations'' have been defined in~\cite{Wiseman03}
as those performed by Alice and Bob in their laboratory after having
measured the number of particles.

Verstraete and Cirac pointed out that the presence of SSR gives rise to a 
new resource which has to be quantified. They have proposed to replace 
the quantity $E_p$ with the {\em SSR-entanglement of formation}. This is 
defined as 
$$
E_f^{(SSR)} [\ket{\psi_N}]  = \min_{p_n,\psi_n} \sum_n p_n E_M[\psi_n]
$$ 
where the minimization is performed over all those decomposition of 
the density matrix where the $|\psi \rangle_n$ are eigenstates of the total number of 
particles~\cite{shuchprl,shuch}.

\subsection{Entanglement for Bosons}

The quantification and classification of boson entanglement is very close
in spirit to that of Fermions as described in Section \ref{ferment}.
We therefore will only emphasize the marking differences for
bosonic entanglement.

In the bosonic case the matrix $\omega$ in Eq.(\ref{gen-ferm-state}) is 
symmetric under permutations of the particle numbers. Consequently, for any 
two-particle state of indistinguishable bosons, $\omega$ can be diagonalized 
by means of unitary transformations of the single particle basis. This leads 
to the Schmidt decomposition for bosons~\cite{Eckert02}.
An curious feature distinguishing this case from the entanglement measures 
of distinguishable particles is that the Schmidt decomposition is not 
unique. In fact, any two equal Schmidt coefficients 
admit for a unitary transformation of the two corresponding basis states, such that 
the superposition of the two doubly occupied states can be written as a 
symmetrized state of two orthogonal states~\cite{Lizeng01,Ghirardi05}.
This is the reason why it is not directly the Schmidt rank, but rather the 
reduced Schmidt rank - obtained after having removed all double
degeneracies of the Schmidt decomposition - that 
determines whether or not a state is entangled. 
This non-uniqueness of the Schmidt rank is also responsible for
the ambiguity of the von Neumann entropy or other purity measures
of the single particle reduced density matrix as an entanglement
measure for Bosons~\cite{Ghirardi05}.

With $z_i$ being the Schmidt 
coefficients with degeneracy $g_i$, the reduced Schmidt rank is at most 
$\frac{g_i}{2}+2\left\{\frac{g_i}{2}\right\}$, where $\{.\}$ denotes the 
non-integer part. As a consequence, a Schmidt rank larger than two implies 
the presence of entanglement. Schmidt rank $2$ with degenerate Schmidt 
coefficients can be written as a symmetrized product of orthogonal
states and consequently is disentangled~\cite{Ghirardi05}.
This feature is also present in the $N$-boson case, where in presence of 
up to $N$-fold degenerate Schmidt coefficients the corresponding state can 
be rewritten as a symmetrization of a product. 

For bipartite systems $\omega$ has full Schmidt rank if $\det \omega\neq 0$. 
A Schmidt rank $1$ can be verified by the same contraction technique 
described for the Fermion case in the previous section, where the Pfaffian 
must be replaced by the determinant. This applies to both the bipartite and the
multipartite case~\cite{Eckert02}.

${}$
\newpage
${}$
\newpage

\chapter{Entanglement in condensed matter systems}

Traditionally many-body systems have been studied by looking for 
example at their response 
to external perturbations, various order parameters and excitation spectrum. 
The study of the ground state of many-body systems with methods developed in 
quantum information might unveil new insight. In this Section we classify 
the properties of the ground state of a many-body system according to its 
entanglement. We first concentrate on spin systems. 
Spin variables constitute a good example of distinguishable objects, 
for which the problem of entanglement quantification is most developed. 
Various aspects mainly of pairwise entanglement will be discussed with 
some short reference on the properties of bipartite entanglement - as
the block entropy - and a comment on the localizable entanglement.
Then will we change focus onto itinerant fermion systems. 

\section{Model systems}

The model Hamiltonian for a set of localized spins interacting via
nearest neighbor exchange coupling in a $d$-dimensional lattice 
can be written as    
\begin{eqnarray}
  {\cal {H}}(\gamma,\Delta,h^z/J)& = & J \sum_{\langle i,j \rangle} 
                              \left[\bigfrac{1+\gamma}{2} S_i^x S_{j}^x +
                              \bigfrac{1-\gamma}{2} S_i^y S_{j}^y 
                              +  \Delta S_i^z S_{j}^z\right] 
                              \nonumber \\
                             && - h^z \sum_{i} S_i^z \; .
\label{general-spin}
\end{eqnarray}
In the previous expression $i,j$ are lattice points, 
$\langle \cdot \rangle$ constraints the sum over nearest neighbors 
and $S_i^\alpha$ ($\alpha=x,y,z$) are spin-$1/2$ operators. 
The nomenclature of the various model deriving from Eq.(\ref{general-spin}) 
is shown in table~\ref{tablemod}.
\begin{table}
\begin{center}
\begin{tabular}{c|r|r}
Model & $\gamma$ & $\Delta$ \\
\hline
\hline
$XX$     &  0         &   0      \\
$XY$     &  $\ne$ 0   &   0      \\
$XXX$    &        0   &   1      \\
$XXZ$    &        0   &   $\ne$ 0\\
$XYZ$    &  $\ne$ 0   &   $\ne$ 0\\
Ising  &        1   &         0
\end{tabular}
\end{center}
\caption{Nomenclature of the various models deriving from Eq.(\ref{general-spin})}
\label{tablemod}
\end{table}
A positive (negative) exchange coupling $J$ favors 
anti-ferromagnetic (ferromagnetic) ordering in the $xy-$plane. 
The parameters $\gamma$ and $\Delta$ account for the anisotropy in 
the exchange coupling, $h$ is the transverse magnetic field.

The ground state of Eq.(\ref{general-spin}) is in general entangled, but  
for any value of the coupling constants $\gamma$ and 
$\Delta$, $J>0$ a value $h_{\rm f}$ for the magnetic field exists
in $d=1,2$, where the ground state is factorized~\cite{Kurmann82,Firenze05}. 
The so called factorizing field $h_{\rm f}$ is given by 
$$
h_{\rm f} = \frac{z}{2}J \sqrt{(1+\Delta)^2 - (\gamma/2)^2}
$$ 
where $z$ is the coordination number of the lattice. 
Note that the result for the factorizing field is rigorous irrespective 
the integrability of the Hamiltonian.

In $d=1$ the model is exactly solvable in several important cases. 
In the next two paragraphs we illustrate some of the results obtained 
for the exactly solvable transverse XY model ($\Delta =0$ and $0\leq\gamma\leq 1$).
The quantum Ising model is a special case corresponding to $\gamma=1$ 
while the (isotropic) $XX$-model is obtained for $\gamma = 0$. 
In the isotropic case the model possesses an additional symmetry, 
resulting in the conservation of the magnetization along the $z$-axis. 
For any value of the anisotropy the model can be 
solved exactly~\cite{LIEB,PFEUTY,McCOY}   
by a Jordan-Wigner- and a successive Bogoliubov transformation.
The Lipkin-Meshkov-Glick model~\cite{LMG1} which also will appear 
in the discussion, emerges from the transverse XY-models when the 
range of spin-exchange, or equivalently the connectivity, is set to infinity.

The properties of the Hamiltonian are governed by 
the dimensionless coupling constant $\lambda=J/2h$.
In the interval  $0<\gamma\le 1$, the system undergoes a second order 
quantum phase transition at the critical value $\lambda_c=1$~\cite{Sachdev99,Takahashibook99}. 
The order parameter is the in-plane magnetization (e.g. in $x$-direction: 
$\langle S^x\rangle $), which is different 
from zero for $\lambda >1$. The magnetization along the $z$-direction,
$\langle S^z\rangle $, is different from zero for any value of $\lambda$ 
with singular behavior of its first derivative at the transition. 
This is reflected also in the singularity present in the second derivative of 
the ground state energy with respect to $\lambda$. In the whole interval
$0<\gamma\le 1$ the transition belongs to the Ising universality class. 
For $\gamma=0$ the quantum phase transition is of the 
Berezinskii-Kosterlitz-Thouless type.

\section{Bipartite entanglement and quantum phase transitions}
\label{ground-critical}

Many scientific investigations have been devoted to the study of entanglement close to 
quantum phase transition (QPT). In contrast to a standard thermodynamic phase transition,
a QPT is a phenomenon that occurs at zero temperature. 
Its essence consists in a significant qualitative change of the ground state of a model 
Hamiltonian, which is induced by the change of an external parameter 
or coupling constant~\cite{Sachdev99}. The main idea is that this drastic 
change of the ground state should be accompanied by characteristic 
entanglement patterns.
Similar to standard phase transition, also a quantum critical point 
is characterized by a diverging correlation length $\xi$, 
which is responsible for the singular behavior of different 
physical observables.
The critical properties in the entanglement we are 
going to summarize below admit a screening of the qualitative change 
of the state of the system experiencing a quantum phase transition. 
In order to avoid possible confusion, it is worth to stress that the 
study of entanglement close to quantum critical points is not supposed to provide 
new insight into the scaling theory of quantum phase transitions;
rather, it may be useful for a deeper characterization of 
the ground state wave function of the many-body system undergoing a 
phase change.
In the following, we shine some light on the behavior of the pairwise entanglement, 
with a subsequent glance at the block entropy~\footnote{QPTs were also studied  
by looking at quantum fidelity~\cite{cozzini1,cozzini2} or the effect of 
single bit operations~\cite{giampaolo1,giampaolo2}}.

Pairwise entanglement close to quantum phase transitions was originally 
analyzed in~\cite{Osborne02,OstNat} for the quantum XY model in transverse
magnetic field in one spatial dimension. 
For the quantum Ising chain, 
the concurrence tends to zero for  $\lambda\gg 1$ and $\lambda \ll 1$, 
where the ground state of the system 
is fully polarized along the $x$-axis and the $z$-axis, respectively. 
Whereas the full polarization in z-direction for small $\lambda$
is guaranteed by the large magnetic field in z-direction, this is not 
the case in the opposite limit for generic values of $\gamma$, and the
polarization in this case is an effect due to symmetry breaking.
A particularly surprising observation is the short range of the concurrence,
in particular at the critical point, notwithstanding the diverging
range of two-point spin correlations: 
the concurrence is zero unless the two sites 
are at most next-nearest neighbors.
This short range in the pairwise entanglement 
is observed in many different models~\cite{Syl,Syl2,AFO07}, also in
higher spatial dimensions. It hence seems to be a generic
feature rather than a curious exception. 
This indicates that pairwise entanglement
typically plays a secondary role in Cavour of multipartite entanglement,
as far as quantum phase transitions are concerned.  

\begin{figure}\centering
\includegraphics[width=8cm]{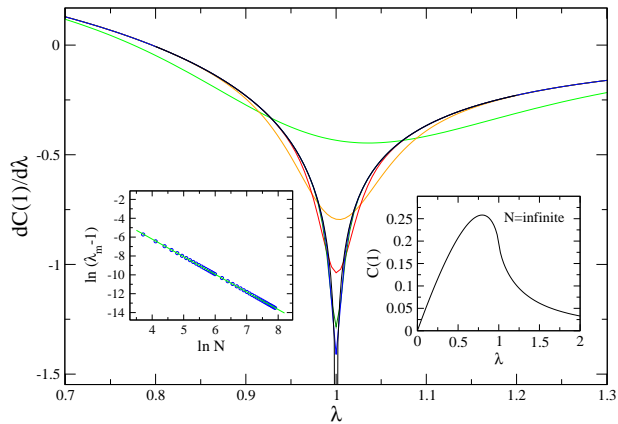}
\vspace*{0.3cm}
\caption{
The change in the ground state wave-function in the critical region 
is analyzed considering the derivative of the nearest neighbor concurrence 
as a function of the reduced coupling strength. Different curves correspond to 
different lattice sizes. On increasing the system size, the minimum gets 
more pronounced. 
Also the position of the minimum changes and tends as $N^{-1.86}$  
(see the left side inset) towards the critical point where for an 
infinite system a logarithmic divergence 
is present. The right hand side inset shows the behavior of the concurrence 
itself for an infinite system. The maximum is not related to the critical 
properties of the Ising model. [From \protect\cite{OstNat}]}
\label{a-derivative-gs}
\end{figure}

In the Ising case, the concurrence is a smooth function 
of the coupling with its maximum well separated from the critical point 
(see the right inset of Fig.\ref{a-derivative-gs}).
In contrast, the convex roof of the von Neumann entropy shows a
pronounced cusp at the critical point~\cite{Osborne02}.
The critical properties of the ground state are instead well reflected in the 
derivatives of the concurrence as a function of $\lambda$. 
The results for systems of different size (including the thermodynamic limit) 
are shown in Fig.\ref{a-derivative-gs}.
For the infinite chain $\partial_\lambda C(1) $ 
diverges on approaching the critical value as 
\begin{equation}
\partial_\lambda C(1) \sim \frac{8}{3 \pi^2} \ln|\lambda-\lambda_c| \;\; .
\label{nn-concurrence}
\end{equation}      				
For finite system size, the precursors of the critical behavior 
can be analyzed by means of finite size scaling. 
In agreement with the scaling hypothesis, 
the concurrence depends only on the combination 
$N^{1/\nu}(\lambda-\lambda_m)$ in the critical region,   
with critical exponent $\nu=1$; $\lambda_m$ is here the position of the minimum
(see the left inset of Fig.\ref{a-derivative-gs}). 
In the case of log divergence the scaling ansatz has to be adapted for taking 
care of the critical characteristic log divergence in the quantum Ising
universality class
\begin{eqnarray}
\partial_\lambda C(1)(N,\lambda) &-&
\partial_\lambda C(1)(N,\lambda_0) \nonumber \\ 
&\sim&
Q[N^{1/\nu}\delta_m(\lambda)] - Q[N^{1/\nu}\delta_m(\lambda_0)]
\end{eqnarray}
where $\lambda_0$ is some non critical value, 
$\delta_m (\lambda) = \lambda-\lambda_m$ 
and $Q(x)\sim Q(\infty) \ln x$ (for large x). 
Similar results have been obtained for the $XY$ models in this universality 
class~\cite{OstNat}. 
Although the concurrence describes short-range properties, 
the typical scaling behavior for continuous phase transitions emerges.
The analysis of the finite size scaling in the so called period-$2$ and 
period-$3$ chains, where the exchange coupling varies every second and 
third lattice sites respectively, leads to the same 
scaling laws in the concurrence~\cite{Zhang05}. 

\begin{figure}[ht]
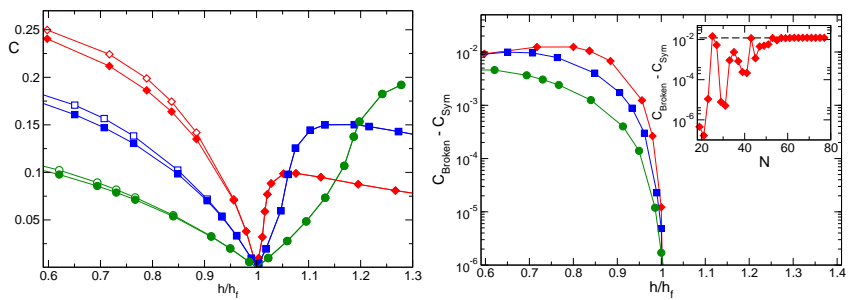

\includegraphics[width=.45\textwidth]{concL200.eps}
\includegraphics[width=.45\textwidth]{concdiffL200.eps}
\caption{{\em Left panel}: The nearest neighbor 
concurrence for a chain of 199 sites
is shown for three different values $\gamma=0.3$, $0.5$ and $0.7$ 
(circles, squares and diamonds). Full symbols give the results for 
the even parity ground state. 
{\em Right panel}: Difference of the n.n. concurrence with and without
broken parity symmetry as a function of the transverse field. 
The maximum relative deviation amounts to around $10\%$;
for sufficiently small $h$, it decreases with $\gamma$.
Inset: Finite size scaling for $\gamma=0.7$ and $h/h_f=0.8$ (diamonds) 
and limiting value (dashed line).[From~\cite{oster06}]}
\label{SBconc}
\end{figure}

Spontaneous symmetry breaking can influence the entanglement in the 
ground state. To see this, it is convenient to introduce the 
thermal ground state in the limit $T\to 0$
\begin{equation}
\rho_0=\frac{1}{2} \left (|gs^o\rangle\langle gs^o| +
|gs^e\rangle\langle gs^e|\right ) = 
\frac{1}{2} \left (|gs^-\rangle\langle gs^-| +
|gs^+\rangle\langle gs^+|\right )\;\; .
\end{equation}
The symmetry broken states $gs^+$ and $gs^-$, which
give the correct order parameter of the model, are superpositions
of the degenerate parity eigenstates $gs^o$ and $gs^e$.
This is the essence of the symmetry breaking in the transverse XY models.
Being convex, the concurrence in 
$gs^{\pm}$ will be larger than for $gs^{o/e}$~\cite{oster06}. 
The opposite is true for the entropy of entanglement 
(see Ref.~\cite{Osborne02} for the single spin von Neumann entropy). 
It was shown, however, that the effect of the spontaneous parity symmetry 
breaking does not affect the concurrence in the ground state if it
coincides with $C^{I}$, Eq.(\ref{e.C'r}): that is, if
the spins are entangled in an anti-ferromagnetic way~\cite{Syl}.
For the quantum Ising model, the concurrence coincides with $C^{I}$ for
all values of the magnetic field, and therefore, the concurrence is unaffected
by the symmetry breaking, the hallmark of the present QPT.
For generic anisotropies $\gamma$ instead,
also the parallel entanglement $C^{II}$ is observed precisely for 
magnetic fields larger than the factorizing field~\cite{OsterlohSPIE};
this interval includes the critical point~\footnote{To see this, it is enough to realize that
a product state of single spin states is a parity eigenstate
only if all spins point up or down.}. 
An interpretation of this is that for the Ising universality class,
the concurrence is insensitive to the symmetry breaking close
to the critical point, and hence won't play a relevant role in driving
this transition. This changes at $\gamma=0$, where the concurrence
indeed shows an infinite range. 
Below the critical field, the concurrence is enhanced by the parity 
symmetry breaking~\cite{oster06}, as shown in Fig.~\ref{SBconc}.

Several works were devoted to an entanglement analysis 
close to this factorizing field. 
The point at which the state of the system becomes separable 
comes with an exchange of parallel and anti-parallel sector 
in the ground state concurrence (see  Eqs.(\ref{e.C'r}) and (\ref{e.C''r})). 
Furthermore, it is observed that the range of the concurrence diverges close
to the factorizing field, in contrast to the typically encountered
short range of the concurrence at the critical point. 
There, the range $R$ is taken as the distance of two qubits, beyond
which the concurrence is zero.
For the $XY$ model the range was found to diverge close to the factorizing 
field as~\cite{Amico06}
\begin{equation}
R\propto \left (\ln \frac{1-\gamma}{1+\gamma}\right )^{-1} \ln |\lambda^{-1}-
\lambda_f^{-1}|^{-1}
\end{equation}     
The divergence of $R$ suggests, as a consequence of the monogamy of the 
entanglement~\cite{Coffman00,Osborne06}, that the role of pairwise 
entanglement is enhanced while approaching the separable 
point~\cite{Firenze04,Roscilde.jltp,Firenze05}.
Indeed, for the Ising model (i.e. $\gamma=1$), 
the ratio $\tau_2/\tau_1 \rightarrow 1$~\cite{Amico06},
when the magnetic field approaches the factorizing field $h_{\rm f}$.  
For $\gamma\neq 1$ and $h_{\rm f}<h^z < h_c$ 
it was found that $\tau_2/\tau_1$ monotonically increases 
with $h^z\rightarrow h_{\rm f}^+$ and that
$(\tau_2/\tau_1)|_{h_{\rm f}^+}$ increases with $\gamma\rightarrow 1$.
The existence of a factorizing field emerged as a generic feature
of spin chains both for short~\cite{Amico06,Firenze04,Roscilde.jltp} 
and long ranged interactions~\cite{DusVidPRL}; 
in all these cases the range of the two-site entanglement was 
observed to diverge. 

The section shall be closed with a brief remark concerning the localizable entanglement.
From its definition and the fact that it is bound from below by the largest two-point
correlation function, it has been presented as a curiosity that an infinite range of entanglement
can be present although the correlation functions have finite range.
As an example for such behavior the AKLT model, a spin-1 model which exhibits a Haldane gap,
has been presented~\cite{popp05}. The infinite range of the localizable entanglement means that
by virtue of suitable local transformations, different classes of entanglement can be accumulated
on two sites (see also~\cite{POpescu92} in this context).
At the same time, however, also the classical correlations are accumulated and hence would also 
these {\em localizable classical correlations} have an infinite range.
The curiosity of having infinite range entanglement but only finite range correlations
is hence only due to looking at qualitatively different and not comparable quantities.

\section{Dynamics of entanglement}
The interest in studying the properties of entanglement in many-body
systems has been directed also to the understanding of its 
dynamical behavior. Entanglement dynamics has been  studied from 
different perspectives. In a spirit similar to the study of propagation 
of excitations in condensed matter systems, several works analyzed the 
propagation of entanglement starting from a given initial  state where 
the entanglement has been  created in a given portion of  the many-body 
system. One can imagine for example to initialize a spin chain such that 
all the spins are pointing upwards except for two neighbor spins which are 
entangled. Due to the exchange interaction, the initially 
localized entanglement will spread. This propagation 
may be ballistic in clean systems or diffusive if some weak disorder is 
present. Entanglement localization and chaotic behavior could 
also be observed. An alternative approach is to start with the ground state 
of a Hamiltonian $H_0$ and then let the Hamiltonian change in time. 

Since we are dealing with interacting systems, entanglement can be 
generated or it can change its characteristics during the dynamical 
evolution. Besides the interest in their own, attention to these questions has 
been also motivated by the potential use of one-dimensional spin systems 
as quantum channels~\cite{bose03}.  
Another important aspect of entanglement dynamics
is the possibility to generate entangled states with certain desired
properties by those interactions that are naturally present in a many-body 
system. This generalizes the setup where a 
Bell state is created by letting two qubits interact for a fixed time by 
means of an exchange coupling of the $XX$ type. In the same spirit one can 
think of generating three-bit entangled GHZ or W states 
$\ket{W}\sim \ket{100}+\ket{010}+\ket{001}$ - but also
other multipartite entangled states - by tailoring the 
exchange couplings in spin networks or their quantum optical counterpart
of atoms in an optical trap potential. 
Cluster states are a prominent example for 
genuinely multipartite entangled states, which are generated
by Ising type two-spin interactions from a fully polarized initial 
state~\cite{hein04}.

The most simple situation, which we consider first, is the propagation of 
entanglement in the one-dimensional $XX$-model, i.e. $\gamma=0$ and $\Delta=0$
in Eq.(\ref{general-spin})~\cite{amico04,subrahmayam04}.
This corresponds to free spinless fermions on a discrete lattice. 
Suppose that the initial state of the chain is 
\begin{equation}
	| \Psi_{\pm} (t=0) \rangle \equiv
	\frac{1}{\sqrt{2}}(\sigma^{x}_i \pm \sigma^{x}_j)|0, \dots 0 \rangle \, ,
\label{wavefunction}
\end{equation}
namely all the spin are in a fully 
polarized state except the two at positions $i$ and $j$, which are prepared in one of the 
Bell states $|\psi_{\pm}\rangle = 2^{-1/2} (|01\rangle \pm |10 \rangle)$.
In this case the problem has a simple analytical solution. 
The total magnetization is conserved, and the evolution 
is confined to the sector with only a single spin pointing up
(single-magnon states). 
The state of a periodic chain at later times is
\begin{equation}
	|\Psi_{\pm} (t) \rangle = \sum_l w^{(i,j)}_{\pm,l}(t) | {\boldsymbol l} \rangle
\label{waveevolv}
\end{equation}
with
\begin{equation}
	w^{(i,j)}_{\pm,l}(t) = \frac{1}{\sqrt{2N}} 
	\sum_{k} \left [ 1 \pm e^{\frac{2 \pi i k }{N}(j-i)} \right ] 
	e^{\frac{2 \pi i k }{N}(i-l)} 
	e^{4iJt \cos \frac{2\pi k}{N}}
\end{equation}
In the thermodynamic limit, $N\rightarrow \infty$, the coefficients can be 
expressed in terms of Bessel functions $J_n(x)$
\begin{equation}
	w^{(i,j)}_{\pm,l}(t) = \frac{1}{\sqrt{2}} \left \{
	J_{i-l}(4J t) \pm \,
 	(-i)^{j-i} J_{j-l}(4Jt) \right \} \; .
\label{bessel}
\end{equation}
The concurrence between two sites located at positions $n$ and $m$  
(initially, only the sites $i$ and $j$ were maximally entangled) is given as  
\begin{equation}
	C^{i,j}_{n,m}(\pm,t) = 
	2 \Bigl | w^{(i,j)}_{\pm,n}(t) w^{(i,j)\star}_{\pm,m}(t) \Bigr | \; .
\label{concnm}
\end{equation}
\begin{figure}
	\vspace*{2mm}
	\includegraphics[width=.87\linewidth]{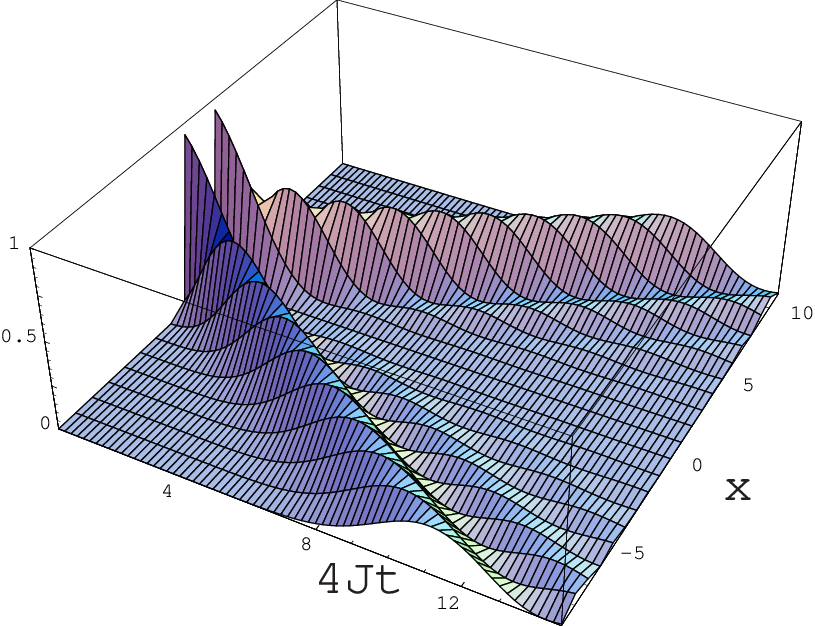}
	\caption{Concurrence between sites $n=-x, m=x$, symmetrically
	placed with respect to the sites $i=-1$ and $j=1$, where the singlet was initially created. 
	[From \protect\cite{amico04}].} 
\label{treconc}
\end{figure}
and is shown in Fig.\ref{treconc} for sites 
which are symmetrically arranged around the initial position of the Bell state 
$|\psi_{\pm}\rangle$. The Hamiltonian time evolution of the $XX$ model 
leads to a propagation of the single flipped spin through the chain. The speed of the 
propagation is the spin wave velocity in the chain. The information exchange 
or entanglement propagation over a distance of $d$ lattice constants
approximately takes the time $t \sim \hbar d/J$. 
The result is independent of the external field $h^z$, since the magnetization
in $z$-direction is a constant of the motion.

\begin{center}
\begin{figure}
\includegraphics[width=1\linewidth]{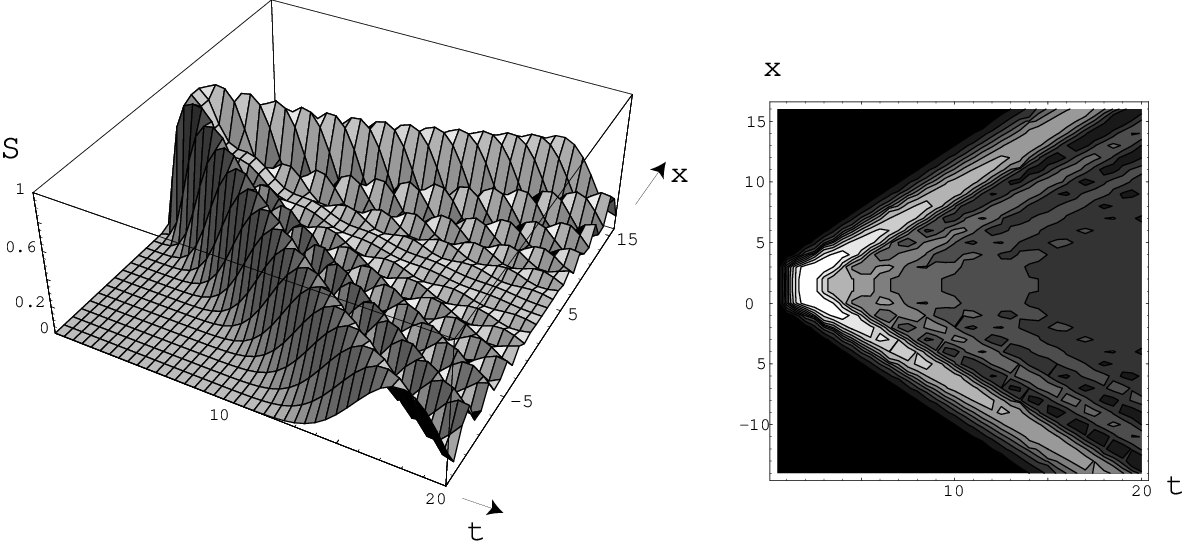}
	\caption{The two-site entropy $S^{(2)}_{(-n,n+1)}(t)$ for pairs of sites
	symmetrically displaced with respect to the initial singlet
	position $(i,j)= (0,1)$. The time is given in units of the 
        exchange coupling $4J$ and the space coordinate $x$ in units of the lattice 
        constant.
	[From \protect\cite{amico04}].}
\label{entro}
\end{figure}
\end{center}
This {\em entanglement wave} is also the main feature in the behavior 
of the von Neumann entropy $S^{(2)}_{n,m}$ 
of the two sites $(n,m)$ (see Fig.\ref{entro}). 

\begin{center}
\begin{figure}
\includegraphics[width=.4\linewidth,angle=-90]{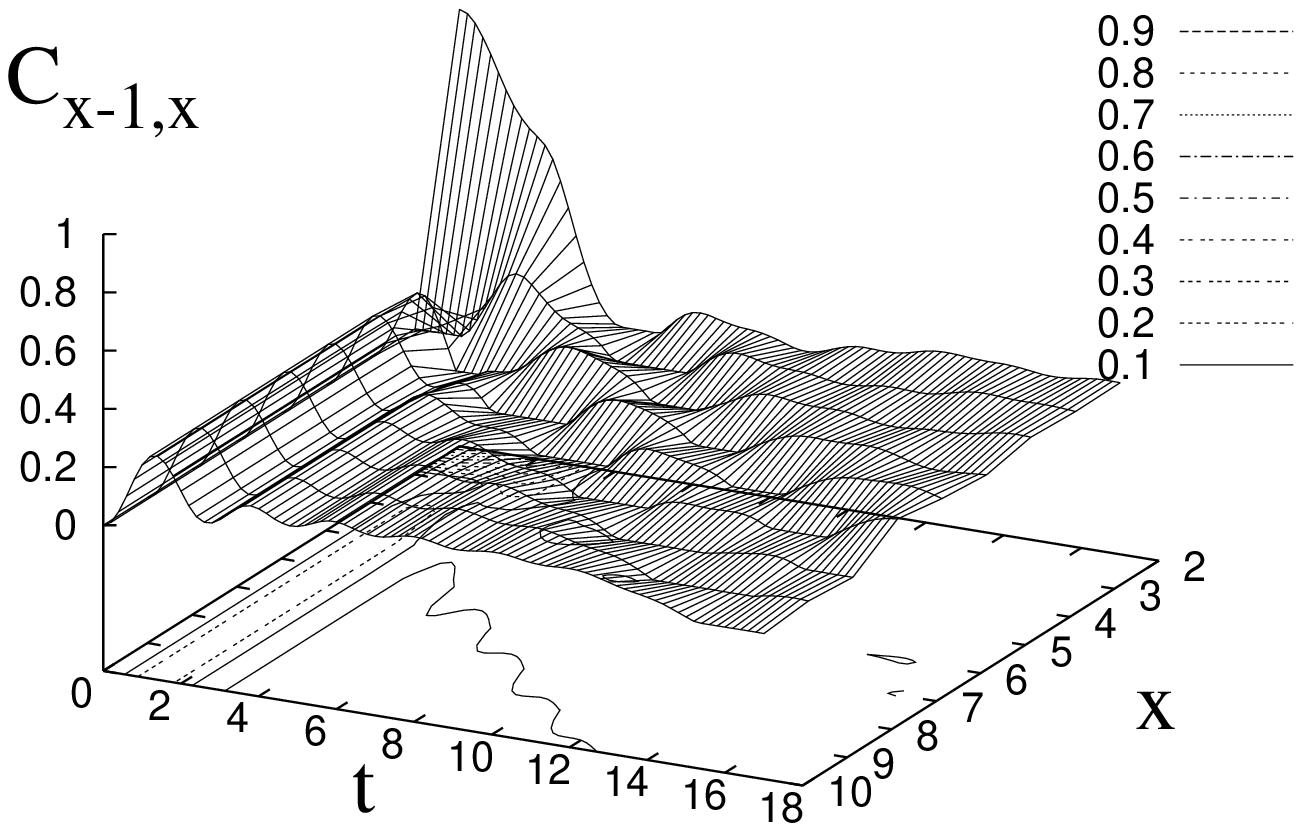}
\includegraphics[width=.4\linewidth,angle=-90]{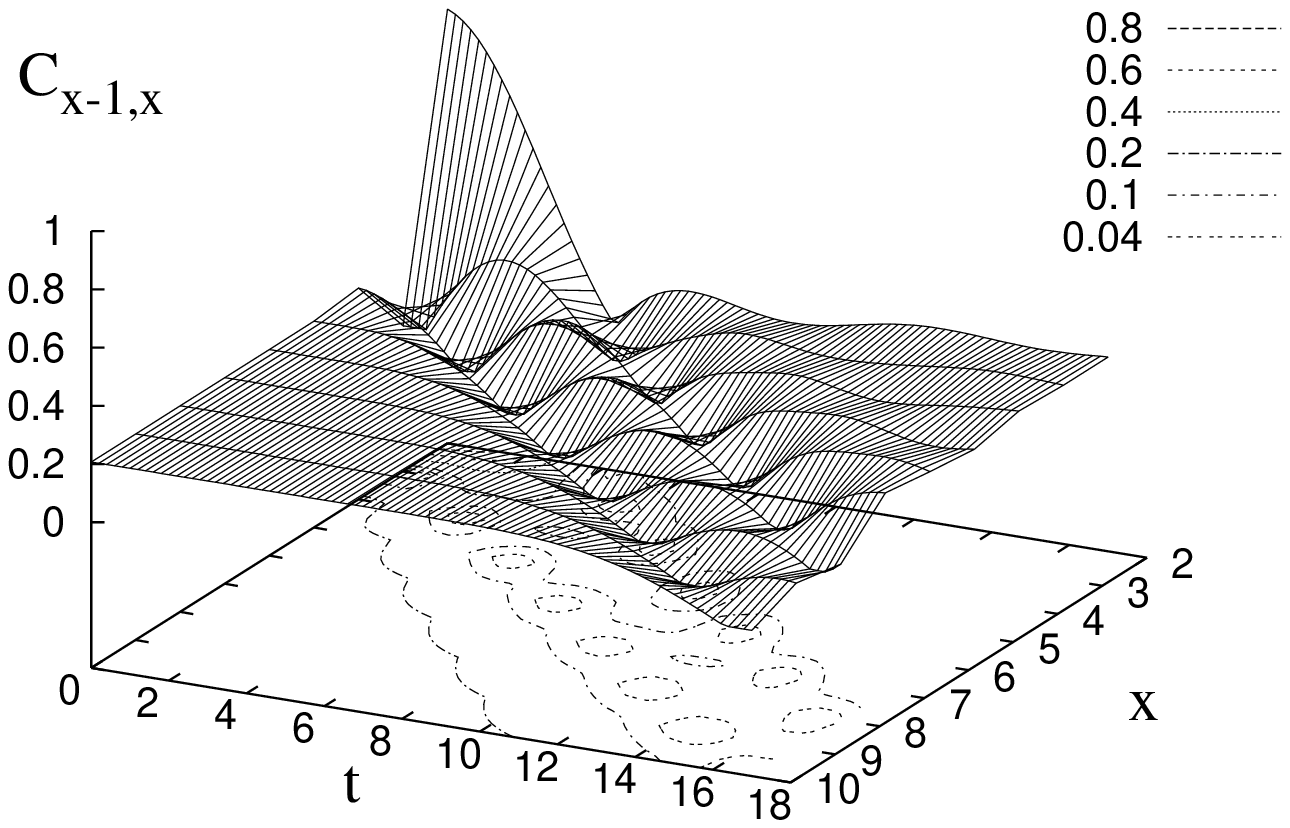}
	\caption{The entanglement wave for the nearest neighbor concurrence is
shown here for the Ising model and $\lambda=0.5$.
The initial state is a maximally entangled state of nearest neighbors
on top of the fully polarized vacuum state (left) and the 
ground state (right) respectively. 
Whereas the propagation velocity is unaffected by the
initial state, the entanglement wave reduces the background entanglement
in the ground state of the chain (right).[From~\cite{amico04}]}
\label{C1dyn}
\end{figure}
\end{center}

\begin{center}
\begin{figure}
\includegraphics[width=.75\linewidth,angle=-90]{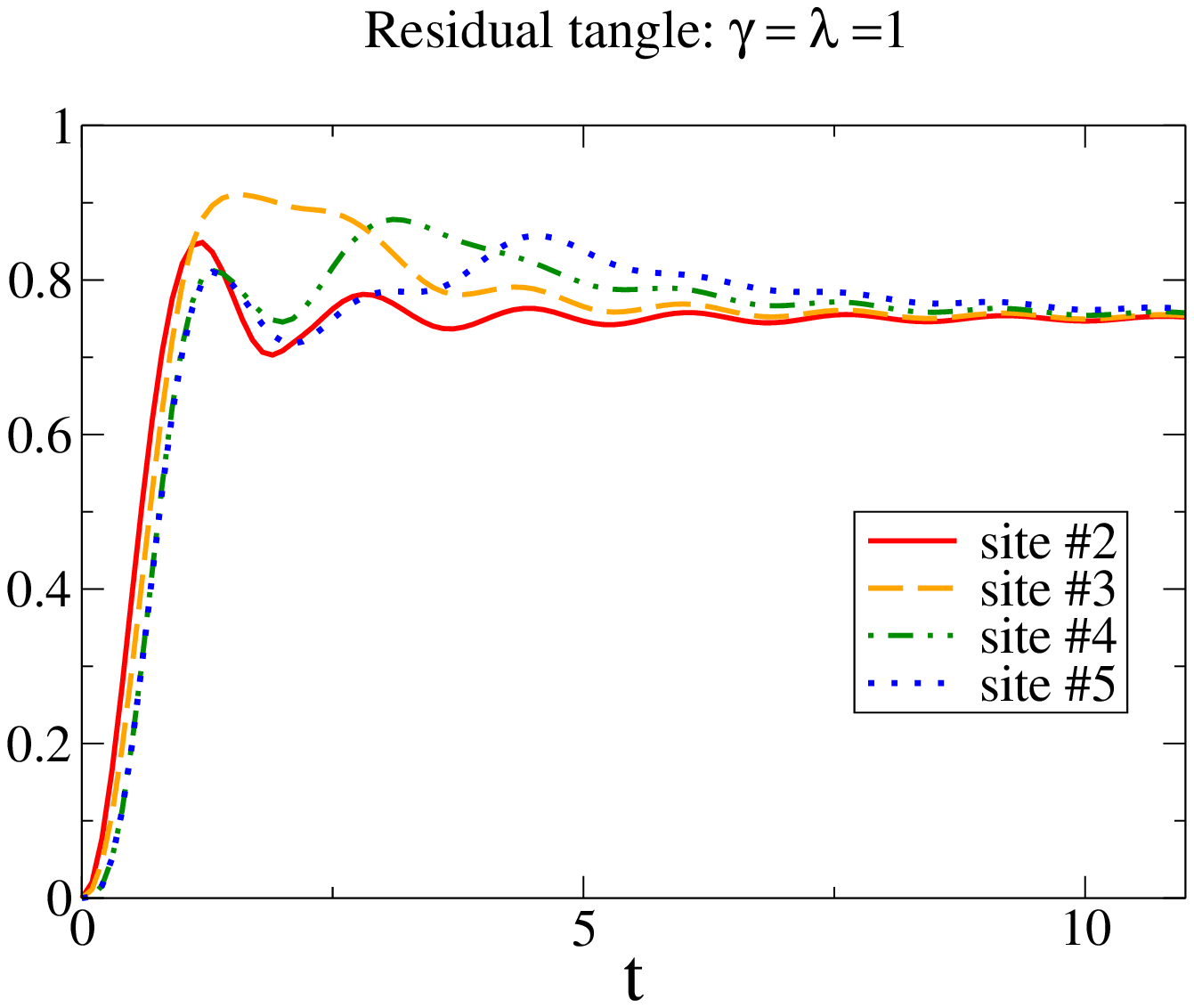}
	\caption{The residual tangle in case of the Ising model and critical
coupling created from the fully polarized vacuum. 
Multipartite entanglement is clearly dominant over pairwise entanglement.[From~\cite{OsterlohSPIE}]}
\label{CKW-dyn}
\end{figure}
\end{center}

Interesting additional features appear in the quantum $XY$ model, i.e. for
$\gamma \ne 0$. 
In this case the magnetization is no longer a constant of the motion 
(two spins can be flipped simultaneously). The calculations were done 
analytically~\cite{amico04} resorting on the exact calculation of
correlation functions out of equilibrium~\cite{Amico-Osterloh}. 
The most notable difference in the two-site entanglement 
is an entanglement production from the fully polarized vacuum state.
This occurs uniformly along the 
chain and is superposed onto the entanglement wave discussed 
before (see left panel of Fig.\ref{C1dyn}). 
The propagation  velocity of the entanglement is almost 
independent on the anisotropy parameter $\gamma$ and is in well
agreement with the sound velocity of the system~\cite{Stolze95}. 
What is strongly dependent on $\gamma$ is the damping coefficient 
of the entanglement wave: as the anisotropy approaches the 
Ising point $\gamma=1$, the wave is strongly damped and vanishes 
already after few time units ($\sim J^{-1}$) at the critical coupling. 
The Hamiltonian dynamics primarily generates multipartite entanglement
in the chain; this can be seen from the 
residual tangle~\cite{Coffman00,Osborne06}, which is a measure for 
entanglement of not only pairs of spins. It is seen from Fig.~\ref{CKW-dyn}
that the major part of the entanglement in the chain
is indeed of multipartite origin.
This should be expected rather than surprising on the background
that multipartite entangled cluster states are created from
a fully polarized state by means of a two-spin Ising-type interactions.  

\section{Entanglement in second quantized systems}
The theory of entanglement for indistinguishable particles 
conceptually differs from that for distinguishable constituents;
the only factoring state of identical particles 
is that of bosons all being in the very same state.

The major part of the physical applications still concentrates 
on those measures applicable for the entanglement encoded in 
a certain choice of - distinguishable - 
quantum numbers (see \ref{indistparts}).
This analysis captures, of course, a partial aspect 
of the entanglement encoded in these system; alas,
some marking peculiarities due to the indistinguishability of the particles
are ignored in this approach.
We will report on the established results from
these studies and refer to section~\ref{indistparts} for 
their relation to fermionic entanglement.

\subsection{Free fermions}
\label{sec-noninteracting}

The site-based entanglement of spin degrees of freedom
through the Jordan-Wigner transformation of spinless fermions has been exploited for 
calculating the concurrence of nearest neighbor sites and the single site 
von Neumann entropy (see Section \ref{indistparts}) for the one-dimensional 
tight-binding model in presence of a chemical potential~\cite{ZanardiPRA02}. 
This model is equivalent to the isotropic $XX$ model in a transverse 
magnetic field.
In this specific case, no double occupancy can occur and the concept of
entanglement coincides with that for spins 1/2.
It was found that the nearest neighbor concurrence of the ground state at 
$T=0$ is maximal at  half filling\footnote{Due to particle-hole symmetry, 
the concurrence is symmetric around half-filling}.

The continuous limit of the tight-binding fermion model is 
the ideal Fermi gas. In this system, the spin entanglement
between two distant particles has been studied in~\cite{Vedral03}.
There, depending on the dimensionality, the pairwise spin-entanglement
of two fermions has been found to decrease with their distance
defining a finite range $R$ of the concurrence.
The two spin reduced density matrix is
\beq
\rho_{12}=\bigfrac{1}{4-2f^2}
\Matrix{cccc}{1-f^2&0&0&0\\0&1&-f^2&0\\0&-f^2&1&0\\0&0&0&1-f^2} 
\eeq
where $f(x)=d \frac{J_1(x)}{x}$ with $d\in\{2,3\}$ being the space dimension 
and $J_1$ the (spherical for $d$=3) Bessel function of the 
first kind~\cite{Vedral03,Oh04}.
This density matrix is entangled if $f^2\geq 1/2$.
As a consequence, 
there is spin entanglement for two fermions closer than 
$d_0\approx 0.65\frac{\pi}{k_f}$ for $d=3$
and $d_0\approx 0.55\frac{\pi}{k_f}$ for $d=2$
($k_f$ is the Fermi momentum).
A finite temperature tends to diminish slightly the range of 
pairwise spin entanglement~\cite{Oh04}.

It should not be surprising that non-interacting particles
are spin-entangled up to some finite distance.
It is true that the ground state and even an arbitrary thermal state
of non-interacting fermions has vanishing entanglement among the 
particles\footnote{This statement should not be confused with the
non-vanishing {\em entanglement of particles}~\cite{Wiseman03} 
as observed in~\cite{Dowling06}.},
since the corresponding states are (convex combinations of) 
anti-symmetrized product states.
However, disentanglement in momentum space typically leads
to entanglement in coordinate space. A monochromatic plane wave
of a single particle for example corresponds to a $W$ state,
which contains exclusively pairwise 
entanglement in coordinate space for an arbitrary distance of the sites.
Furthermore does a momentum 
cut-off at $k_f$ correspond to a length scale of the order $k_f^{-1}$.

It is interesting that a {\em fuzzy} detection of the particles
in coordinate space increases the entanglement detected by the  
measurement apparatus. 
In ref.~\cite{Cavalcanti05b},
the two-position reduced density matrix defined by
\beq
\rho^{(2)}_{ss',tt'}=\expect{\Psi_{t'}(r')^\dagger\Psi_{t}(r)^\dagger
\Psi_{s'}(r')^{}\Psi_{s}(r)^{}}
\eeq
has been calculated for blurred field operators
\beq
\Psi_{s}(r)^{}:=\int\!\d r' \,\d p\; \psi_s(p) D(r-r')\e^{ipr'}
\eeq
where $D(r-r')=\frac{1}{\sqrt{2\pi}\sigma}\exp{-\frac{|r-r'|}{2\sigma^2}}$
is a Gaussian distribution describing the inaccuracy of the 
position measurement.
This could be understood from the blurred field operators being coherent
sums of local field operators; the entanglement measured by the apparatus 
as described above, is the bipartite entanglement between the two regions
of width $\sigma$ around $r$ and $r'$. This entanglement is larger than
the average of all pairwise contributions out of it due to
the  super-additivity of the entropy/negativity.
An analysis in~\cite{Vedral04} for the three fermion spin density 
matrix revealed that
the state carries entanglement within the W-class~\cite{Duer00}, provided
the three particles are in a region with radius of the order of the 
inverse Fermi momentum; a similar reasoning applies to $n$ fermions
in such a region~\cite{Vedral04,lukens}.

\subsection{$su(2)$ degrees of freedom of interacting fermions} 
\label{supercond}
Itinerant systems, where the focus of interest 
is the entanglement of degrees of freedom forming
a representation of $su(2)$ in terms of the fermionic operators
have been also the subject of investigation.  
This line has been followed in Refs.~\cite{ZanardiPRA02,Shi04,VedralNJP04,fan04,Vedral04}
for analyzing a connection to BCS superconductivity
and the phenomenon of $\eta$-pairing, a possible scenario for 
high $T_c$  superconductivity. 
Such states appear as eigenstates of the Hubbard model~\cite{Yang89}
which carry off diagonal long range order.
A simplified model of BCS-like pairing for spinless fermions
has been studied in~\cite{ZanardiPRA02}.
The concurrence of the two qubits represented by the modes $k$ and $-k$
has been found to be a monotonically increasing function
of the order parameter; it drops to zero significantly before the 
critical temperature is reached, though.

States with off diagonal long range order by virtue of $\eta$-pairing 
are defined from fermionic 
operators $c_{j,\up}$, $c_{j,\down}$ and the fermionic vacuum $\ket{0}$ as
\beqa\label{etapairing}
\eta_j&:=&c_{j,\up}c_{j,\down}\; ;\quad 
\eta:=\sum_{j=1}^L c_{j,\up}c_{j,\down}\\
\ket{\Psi}&=&{\eta^\dagger}^N \ket{0}
\eeqa 
These are symmetric states
and consequently, their concurrence vanishes in the thermodynamic limit
due to the monogamy property of pairwise entanglement of $su(2)$ degrees 
of freedom.
Consequently, a connection to the order parameter of off diagonal long 
range order
\beqa\label{odlro}
{\cal O}_{\eta}&=&\Expect{\Psi}{\eta^\dagger_j\eta^{}_k}{\Psi}\\
&=& \bigfrac{N(L-N)}{L(L-1)} {\longrightarrow} n(1-n) \; \nonumber
\eeqa
(with $N,L\longrightarrow \infty$ and fixed filling fraction $n$)
can not be established,
not even for the rescaled concurrence $C_R$, since
$$
C_R=2{\cal O}\left(1-\sqrt{\bigfrac{(N-1)(L-N-1)}{N(L-N)}}\right)
\longrightarrow 1/L
$$ 
(see also the analysis for the LMG model in 
Section~\ref{ground-critical}).
Nevertheless, the state is entangled, as can be seen from the
entropy of entanglement and the geometric measure of entanglement~\cite{wei03}.
The latter is tightly connected to the relative entropy~\cite{wei04}.
Both have been calculated in Ref.~\cite{VedralNJP04} and clearly indicate
the presence of multipartite entanglement.

\subsection{Hubbard-type models for interacting fermions}
\label{hubbardsent}
An interesting class of interacting fermion models is that
of Hubbard type models. 
The Hubbard model~\cite{HUBBARD-BOOK} is defined by the Hamiltonian
\begin{equation}
                  {\cal H}=  -t \sum_{\langle ij \rangle} 
                     [c_{i,\sigma}^\dagger c_{j,\sigma}+h.c.] +
                    U \sum_i n_{i,\uparrow} n_{i,\downarrow} -\mu N
\label{Hubbardmod}
\end{equation}
where  $c_{i,\sigma}$, $c_{i,\sigma}^\dagger$ are fermionic operators: 
$\{c_{i,\sigma}, c_{j,\sigma'}^\dagger\}=\delta_{i,j}\delta_{\sigma \sigma'}$.  The coupling constant $U$ describes the on-site repulsion, $t$ is the 
hopping amplitude and $\mu$ the chemical potential.

A first study of entanglement in such a system
has appeared in~\cite{gu04} for the one-dimensional
extended Hubbard model for fermions with spin 1/2.
The extension consists in a nearest neighbor density-density coupling $V$.
Due to the conservation of particle number and $z$-projection of the spin,
the local density matrix of the system takes the simple form
\beq\label{rhoExtHubb}
\rho^{(1)}=z\proj{0}+u^+\proj{\up}+u^-\proj{\down}+w\proj{\up\down}
\eeq
independent of the site number $j$ because of translational symmetry.
\begin{figure}[ht]
\begin{center}
\includegraphics[width=0.75\textwidth]{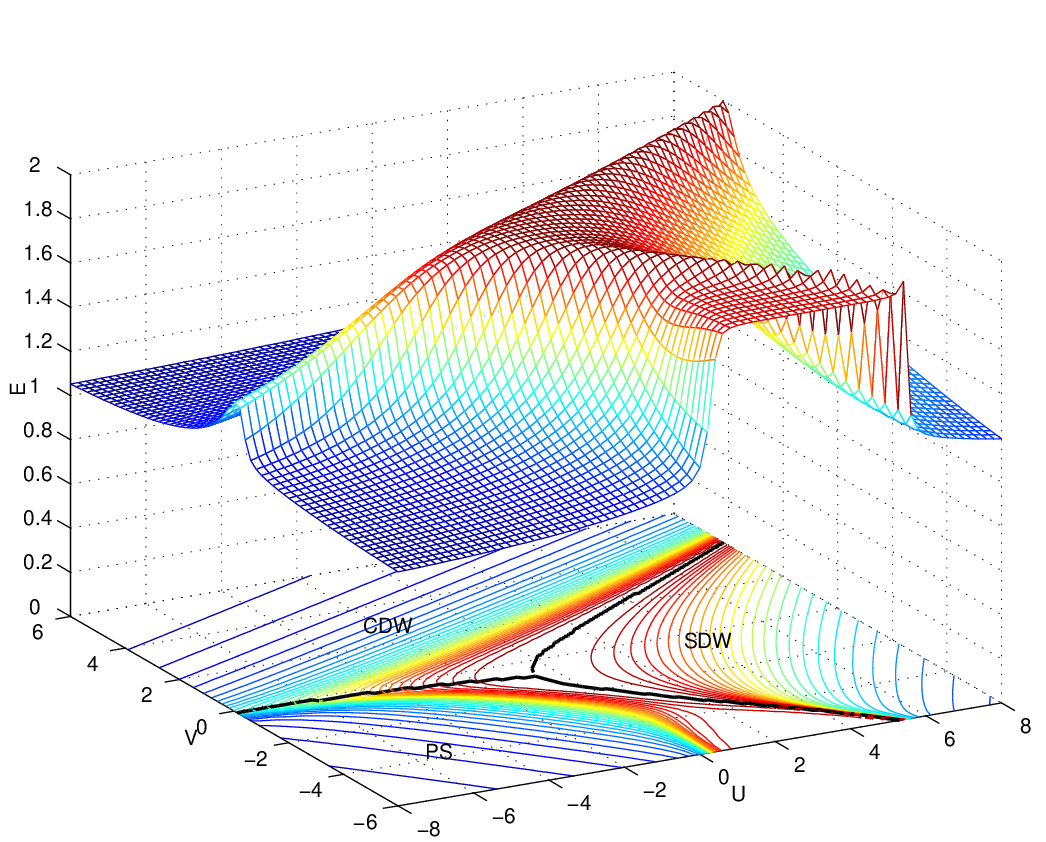}
\caption{
\label{GuDeng1-4}
The local entropy as a function of the on-site coupling $U$ and 
nearest neighbor coupling $V$. The contour plot below includes
the known phase diagram of the model (full black line). Except the 
superconducting phase transition line, the phase diagram is nicely reproduced.
From Ref.~\cite{gu04} }
\end{center}
\end{figure}
The broken translational invariance in the charge density wave phase 
has not been taken into account in this work. This does not affect the 
central result but might affect the entropy within the charge density 
wave phase. Except the superconducting phase, the phase diagram 
at half filling (for $\mu=0$) of this model has been nicely reproduced by 
the contour plot of the local entropy (see Fig.~\ref{GuDeng1-4}),
where the phase transition lines coincide with its crest. 
This happens to be an often encountered feature of local entropies 
- also for spin models - as opposed to the concurrence 
for pairwise entanglement whose maxima in general appear at a certain distance 
to quantum critical points and hence are not associated to the 
quantum phase transition. 
In view of the monogamy of entanglement 
this is evidence for dominant multipartite entanglement 
in the vicinity of quantum phase transitions.

This analysis clearly points out that the local entropy indicates
different phase transitions in different ways, essentially depending on
whether this quantity is sensitive to its order parameter or not.
Due to the $u(1)$ symmetry of the model, the single site reduced 
density matrix is a functional of occupation numbers only.
These operators cannot, however, describe order parameters of
superconductivity or some order parameter of the metal-insulator 
transition.
Indeed, the superconducting phase can be predicted if the entropy of 
entanglement is calculated for a block of spins, instead of for just 
a single site~\cite{DengGu05}. A reduced density matrix of at least
two sites is necessary for being sensitive to superconducting correlations
(see also Ref.~\cite{Legeza06} for a similar result obtained for the 
ionic Hubbard model.)

\begin{figure}[h]
\begin{centering}
\psfrag{u}{\small $u$}
\psfrag{U}{\small $u$}
\psfrag{D}{\small$n$}
\psfrag{n}{\small $n$}
\psfrag{T}{\small $\partial_u\mathcal{S}_{i}$}
\includegraphics[width=0.42\textwidth]{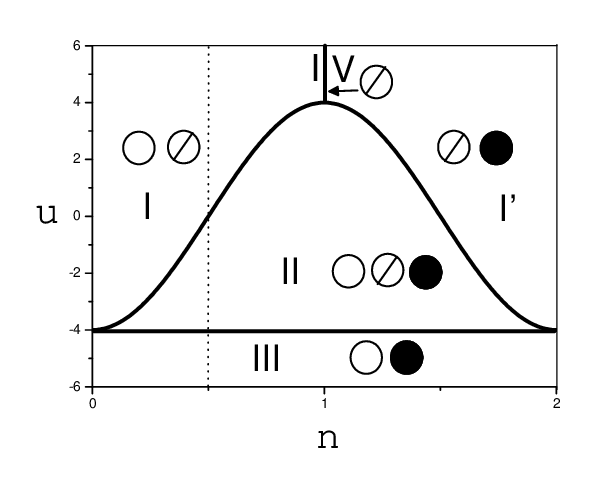}
\hspace*{7mm}
\includegraphics[width=0.34\textwidth, viewport= 60 250 580 825,
clip]{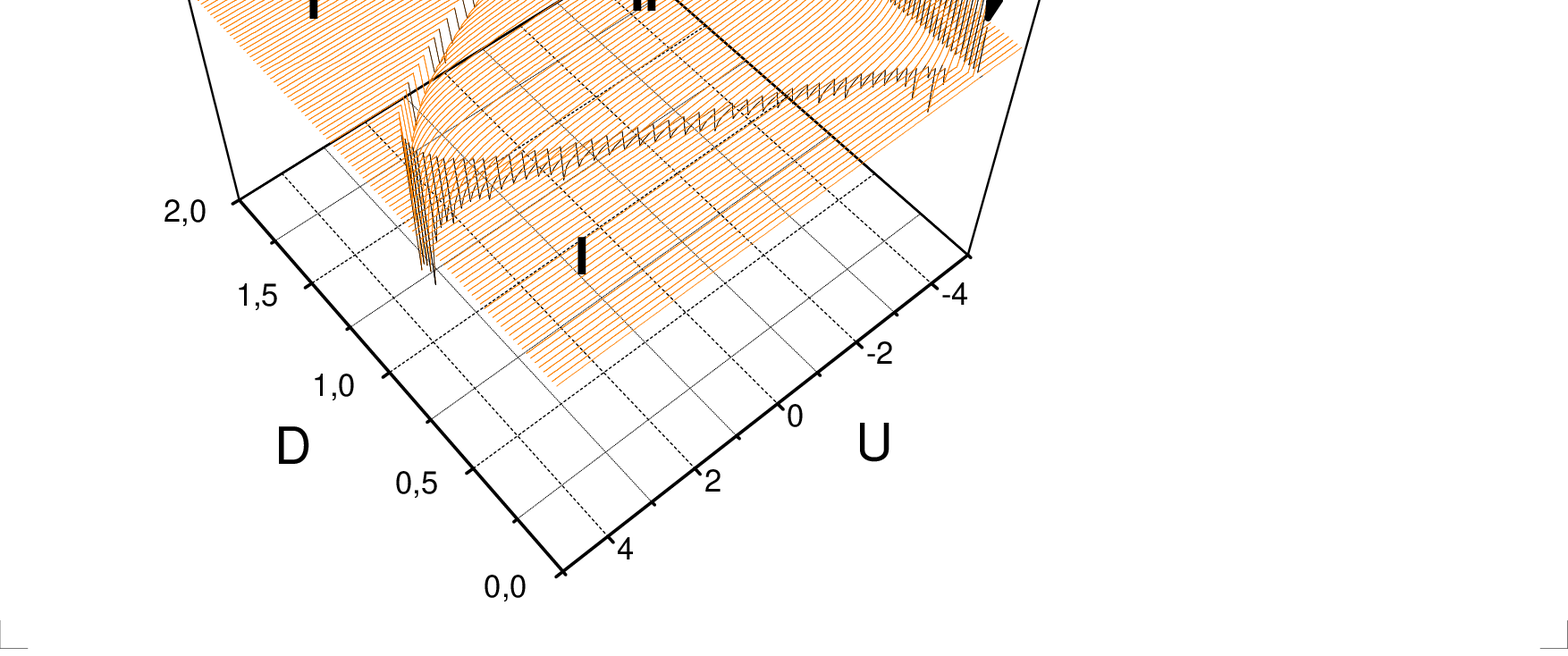}
 \caption{{\em Left panel:} The ground state phase diagram of the
Hirsch model at $x=1$. Empty, slashed and full circles 
indicate the presence of empty, singly and doubly occupied sites,
respectively. 
{\em Right panel:} Except the insulating line IV, 
the phase diagram is nicely reproduced by 
$\partial_u\mathcal{S}_{i}$. [From~\cite{AnfossiGiorda05}]}
 \label{Anfossi1-2}
 \end{centering}
\end{figure}

Another model studied with this respect is the so called
bond-charge extended Hubbard model, also known as the Hirsch model, 
which has originally been proposed
in the context of high $T_c$ superconductivity~\cite{HIRSCH}.
The Hamiltonian is
\begin{equation}
     {\cal H}=U \sum_i^L n_{i,\uparrow} n_{i,\downarrow}
-t[1-x(n_{i,\sigma}+n_{i+1,-\sigma})]c_{i,\sigma}^\dagger c_{i+1,\sigma}+h.c. 
\label{H-fermions}
\end{equation}
For $x=0$ the model (\ref{H-fermions}) coincides with the usual Hubbard model 
(\ref{Hubbardmod}). 
In phases II and III (see Fig.~\ref{Anfossi1-2}) there are superconducting correlations
due to $\eta$-pairing and hence there is
multipartite entanglement, as discussed above~\cite{VedralNJP04}.
For $x=1$, the model is exactly solvable, 
and the entanglement of the model has been analyzed in 
Ref.~\cite{AnfossiGiorda05}.
For general $x$ and $n=1$ see Ref.~\cite{AnfossiBoschi05,Anfossi06}. 
Besides the local entropy of entanglement $S_i$,
also the negativity~\cite{Vidal02}
and the quantum mutual information~\cite{Groisman05} ${\cal I}_{ij}$
have been used and compared for this analysis. 
While $S_i$ measures all (pairwise and multipartite) quantum correlations
involving this specific site, the negativity gives a 
lower bound for the quantum correlation of two specific sites, 
and the mutual information accounts for pairwise quantum and 
classical correlations.
Therefore, this combination of correlation measures opens the possibility
to decide, what type of correlation is relevant at a
quantum phase transition. The local entropy is shown in 
Fig.~\ref{Anfossi1-2}.
The different phases are discriminated by local occupation numbers;
consequently, the entropy $S_i$ bears 
the information on all the phase diagram except the insulating line IV.
This is seen from the plot of $\partial_u S_i$ (with $u\doteq U/t$) as 
a function of $u$ and the filling fraction $n$.
A comparison of first derivatives respect to $y=n,u$
(depending on the phase transition) of all three correlation measures 
reveals common singularities for $\partial_y S_i$ and
$\partial_y {\cal I}_{ij}$ only for the transitions II-III and II-IV; 
furthermore, it is found that the 
range $R$ of the concurrence diverges\cite{Anfossi06} at both transitions. 
These facts allow to judge the transitions II-III and II-IV 
(at $n=1$ and arbitrary $x$) as governed by pairwise entanglement. 
For the transitions II-I  and II-I' instead, multipartite entanglement 
is relevant, accompanied by a finite range of the concurrence. 
A similar behavior was encountered for {\it non-critical} spin models 
where the divergence of $R$ was accompanied by the emergence of a 
fully factorized ground state. Here, $R$ diverges close to QPT.    
 
In order to detect the charge transition II-IV at $n=1$ and any $x$, 
$\partial_x S_i$ has been calculated by means of DMRG~\cite{AnfossiBoschi05}. 
Its singularities allowed to accurately determine the charge gap as a function
of the bond-charge coupling $x$.

We finish this selection of results for itinerant fermion systems 
with the Hubbard model in a magnetic field.
Also here, the local entropy $S_i$ has been looked at in order to analyze 
its entanglement. 
As in the examples before,
$S_i$ indicates the second order phase transitions in terms of
divergences of its derivatives $\partial _h S_i$ and
$\partial_\mu S_i$, respectively. Indeed, it has been demonstrated that
$\partial_h S_i$ and $\partial_\mu S_i$ can be expressed in terms of 
spin and charge susceptibilities~\cite{Larsson05}.
This bridges explicitly between the standard method
in condensed matter physics for studying phase transitions and
the approach from quantum information theory.

Summarizing, the body of work developed so far suggests the 
conclusion that local entropies can detect
QPTs in systems of itinerant fermions, particularly if the transition itself is well predicted
by a mean field approach for local observables of the model. 
In the described cases, translational invariance leads to predictions
independent of the site, the local entropy is calculated for.
In absence of this symmetry, it might prove useful to average over the
sites; the resulting measure is then equivalent to the 
$Q$-measure~\cite{Wallach}.

Though there certainly are transitions with dominant features in
the pairwise entanglement, also here the generic case indicates the 
dominance of multipartite quantum correlations.

\subsection{A remark on entanglement of particles}

There is little work which uses measures for indistinguishable particle
entanglement (see Section~\ref{indistparts}), particularly regarding  
the use of the fermionic concurrence, giving account 
for the possibility of double occupancy (with internal degree of freedom).
The entanglement of particles (see  
Section~\ref{measures-entanglement-particles}) and its difference from
the usual spin entanglement has been worked out in~\cite{Dowling06}; starting with
very small systems as two spinless fermions on four lattice sites and the 
Hubbard dimer, and then for the tight binding model in one 
spatial dimension, the results are compare with previous results for the 
spin entanglement in~\cite{Vedral03}. 

For the Hubbard dimer (a two-site Hubbard model), 
the authors compare with the results for
the entanglement measured by the 
local von Neumann entropy without superselection rule for the local
particle numbers~\cite{ZanardiPRA02}.
Whereas the latter signals decreasing entanglement in the ground state
with increasing $U/t$, the entanglement of particles 
increases~\cite{Dowling06}.
This demonstrates that imposing superselection rules
may lead to qualitatively different behavior of the entanglement.  
Interestingly, an increase with $U/t$ is observed also for the
entanglement of modes (without imposing superselection rules)\footnote{For the extended Hubbard dimer, which is defined as the two site extended Hubbard model, 
see~\cite{DengGu04}}.

We would like to finish this section with the notice of a recent 
proposal of an experiment in order to decide whether 'entanglement'
merely due to the statistics of the indistinguishable particles
can be useful for quantum information processing~\cite{Cavalcanti06}.

\chapter{Multipartite entanglement: quantification and classification}
One conclusion from the preceding chapter is that multipartite
entanglement plays an important role in condensed matter systems,
in particular close to a quantum phase transition.
In this chapter we will present a whole variety of quantities
that have been proposed for the scope of quantifying and eventually
classifying multipartite entanglement.

\section{A zoo of multipartite entanglement measures}
\label{measure-multipartite}

Both the classification of entanglement and its quantification
are at a preliminary stage even for distinguishable particles 
(see however~\cite{Duer00,Miyake02,VerstraeteDMV02,
Briand03,BriandLT03,OS04,OS05,Luque05,mandilara} and references therein).
This uncertainty is responsible for the vast amount of suggested quantities
for the analysis of multipartite entanglement in many-particle systems
and in particularly interesting wave functions.
It has already been mentioned that several quantities applied in the previous 
sections are useful as indicators for multipartite entanglement
when the whole system is in a pure state; then the
cumbersome convex-roof construction is not needed.
For non-degenerate ground states of model Hamiltonians, this requirement
is met at zero temperature $T=0$. 
The entropy of entanglement is an example for such a quantity and several  
works use multipartite measures constructed from and related to
it (see e.g.~\cite{Coffman00,Wallach,Viola03,Scott04,oliveira,Love06}).
These measures are of 'collective' nature - 
in contrast to 'selective' measures -
in the sense that they give 
indication on a global correlation without discerning among the different 
entanglement classes encoded in the state of the system.
The advantage of these measures is that they are easily computed.
Their disadvantage is founded in an ambiguity concerning their choice:
as soon as at least two different entanglement classes are measured,
an infinite variety of inequivalent measures does exist.
As an example, infinitely many proposals could be generated
from~\cite{Scott04,oliveira,oliveira06b,Love06} by substituting
the standard entropy with another mixedness measure
(e.g. the one-parameter family of R\'enyi entropies, etc.).
Since these new measures will induce a different ordering in the space of
entangled states, some results will deviate. The main problem
would then consist in extracting the information about different
entanglement classes from a vast collection of results.
In this context, the authors of Ref.~\cite{facchi06,facchi06a} suggest
the analysis of a distribution of purities for different bipartitions.
If such a distribution is sufficiently regular, 
its average and variance might prove characteristic for the 
global entanglement in the system.
The application of this analysis to the one-dimensional quantum 
Ising model in a transverse field revealed
a well behaved distribution function, whose average and second moment 
are good indicators of the quantum phase transition~\cite{facchi06b}: 
at the quantum critical point both the average and the standard deviation 
exhibit a peak that becomes more pronounced as the number of qubits is 
increased.

Another pragmatic way of quantifying entanglement in a collective way
is represented by the geometric measure of entanglement~\cite{wei03}. 
It quantifies the entanglement of a 
(pure) state through the minimal distance of the state from the set of 
(pure) product states~\cite{RelativeEntropy,wei03}
\begin{equation}
   E_{g}(\Psi) = - \log_2 \max_{\Phi}\mid \langle \Psi | \Phi \rangle \mid ^2
\label{weimeas}
\end{equation}
where the maximum is on all product states $\Phi$. 
As discussed in detail in~\cite{wei03}, the previous definition
is an entanglement monotone. It is zero for separable states 
and rises up to unity for e.g. the maximally entangled n-particle GHZ states. 
The difficult task in its evaluation is the maximization 
over all possible separable states and of course the convex roof extension to mixed states. 

A different approach was pursued in~\cite{guehne05} (see also~\cite{Sharma06})
where different bounds on the average energy of a given system are obtained 
for different types of n-particle quantum correlated states.
A violation of these bounds then implies the presence of 
multipartite entanglement in the system.
The starting point of G\"uhne {\em et al.} are the 
{\em n-separability} and {\em k-producibility} which admit to discrimi\-nate 
particular types of n-particle correlations present in the system. 
A pure state $\mid \psi \rangle$ of a quantum systems of N parties is 
said to be n-separable if it is possible to find a partition of 
the system for which $\mid \psi \rangle = |\phi_1 \rangle |\phi_2 
\rangle \cdots |\phi_n \rangle$.
A pure state $\mid \psi \rangle$ can be 
produced by k-party entanglement (i.e. it is k-producible) if we can write
$\mid \psi \rangle = |\phi_1 \rangle |\phi_2 \rangle \cdots |\phi_m \rangle$
where the $|\phi_i \rangle$ are states of maximally k parties; 
by definition $m \ge N/k$. 
It implies that it is sufficient to generate specific 
k-party entanglement to construct the desired state. 
Both these indicators for multipartite entanglement are collective,
since they are based on the factorizability of a given many particle
state into smaller parts. k-separability and -producibility both
can not discriminate the different k-particle entanglement classes
(as e.g. the k-particle W-states and different k-particle graph states~\cite{hein04}, 
like the GHZ state).  

Another approach pursued is the generalization of the celebrated
concurrence.
For the quantification of pairwise entanglement in
higher dimensional local Hilbert spaces, the concept of
concurrence vectors has been formulated~\cite{Audenaart01,Badziag02}
besides the I-concurrence~\cite{Rungta}; 
the length of the concurrence vector has proved equivalent 
to the I-concurrence~\cite{Wootters01}.
Also for multipartite systems of qubits the concurrence vector
concept has been proposed~\cite{Akhtarshenas05}. In the multipartite setting however
this means to apply the pure state concurrence formula to a mixed
two-site reduced density matrix. It will coincide with the true concurrence
if and only if the eigenbasis of the density matrices accidentally are an optimal
decomposition. Therefore, the concurrence vector in this case is not a vector whose entries
are the concurrences, and it is at least not obvious whether this proposal
is an entanglement monotone. 
 
The {\em n-tangle} is a straightforward extension of the concurrence to multipartite states
as the overlap of the state with its time-reversed~\cite{Wong00}.
It vanishes identically for an odd number of qubits, but 
an entanglement  monotone is obtained for an even number of qubits.
Due to its factorizing structure, it detects
products of even-site entangled states in addition to certain 
genuine multipartite entangled states: it detects the multipartite 
GHZ or cat state, but not for example the four qubit cluster state. 
Therefore, also the n-tangle is a collective measure. 

\section{Measures for genuine multipartite entanglement}

The counterpart to collective entanglement measures
are selective measures for each different multipartite entanglement class separately. 
A first multipartite example beyond pairwise qubit entanglement 
has been derived from the concurrence and the (one-)tangle: 
the 3-tangle~\cite{Coffman00}. It is a measure
for genuine tripartite entanglement that discriminates also from pairwise
entanglement distributed all over the chain, as is the case for the $W$
state $\ket{W}\sim \ket{100}+\ket{010}+\ket{001}$. 
The 3-tangle coincides with the 3-dimensional hyperdeterminant
for two-dimensional local vector spaces~\cite{Miyake03}, i.e. a generalized determinant form
for $3x2$ matrices over $\CC$.
It originated the insight that $SL(2,\CC)$ invariance rather than $SU(2)$ invariance
leads to a classification of genuine multipartite entanglement:
the 3-tangle is indeed the only $SL(2,\CC)^{\otimes 3}$
invariant, and only a single entanglement class with respect to
SLOCC~\cite{Duer00} does exists.
The task of finding SLOCC-class selective measures of qubit entanglement
hence is reduced to finding local $SL(2,\CC)$ invariant operators,
from which global $SL(2,\CC)^{\otimes n}$ invariants can be constructed.
The invariance group should be extended to including the symmetric group
$S_q$ for $q$-partite systems\cite{OS04}\footnote{This amounts to including
the permutation of qubits to the realm of ``local'' operations; this bigger
symmetry group has been termed $SL_{loc}*$ in Ref.~\cite{Dokovic}.}.
The genuine multipartite entangled states belong to the 
{\em non-zero SLOCC class}\footnote{
The complementary {\em zero SLOCC class} is made of those states that 
vanish after - maybe infinitely many - suitable $SL(2,\CC)$ operations;
the non-zero SLOCC class is the complement of the zero SLOCC class.}.

There are standard formalisms in the well developed field of invariant
theory for the construction of {\em all} $SL(2,\CC)^{\otimes n}$ 
invariants (see Ref.~\cite{BriandLT03} and references therein, but also
Ref.~\cite{VerstraeteDM03} for the contraction method with the invariant 
tensor, which for qubits or spins-1/2 is $i\sigma_y$).
However, the construction of a complete set is very cumbersome already for 
five qubits~\cite{Luque05}
and the so constructed invariants will typically be non-zero
also for certain factorizing states, and consequently will they
be collective measures. Furthermore would we desire to
walk in the steps of Wootters' concurrence and to write the invariants
as expectation values of some {\em tangle operator}.
The minimal requirement would then be that 
the measure of a somehow factorizing state should be zero. 
Interestingly, it seems that this 'minimal' requirement
already implies the necessary invariance, at least for 
qubits~\cite{OS04,OS05}.

The special case of factorization
into a single qubit and the remaining (n-1) qubits leads to
the central quality of the local operators we are searching for:
its expectation value should vanish for all single qubit states $\ket{\psi}$
\beq\label{Alin}
\bra{\psi}{\cal O}\ket{\psi}\equiv 0
\eeq
Such an operator can impossibly be linear, and antilinear operators
have to be studied~\cite{OS04}.
The unique antilinear operator satisfying Eq.~(\ref{Alin})
is $\sigma_y{\mathfrak C}$, as used for
the definition of the pure state concurrence, where
${\mathfrak C}$ is the complex conjugation in the eigenbasis of $\sigma_z$. 
It is suggested from the described shortcomings of the n-tangle
as a measure for genuine multipartite entanglement measure~\cite{Wong00}
that at least one further operator besides $\sigma_y$ will be needed.
Such an additional operator is found when the requirement~(\ref{Alin})
is extended to the two-fold copy of single qubit states 
\beq \label{Alin-multi}
\bra{\psi}\bullet\bra{\psi}{\cal O}\ket{\psi}\bullet\ket{\psi}\equiv 0
\eeq
The requirement to being trace-norm orthogonal to 
$\sigma_y\bullet\sigma_y{\mathfrak C}$ leads to the unique
(up to normalization) operator
\beq
\sum_{\mu=0}^3g_\mu\sigma_\mu\bullet\sigma_\mu{\mathfrak C}
=:\sigma_\mu\bullet\sigma^\mu{\mathfrak C}
\eeq
where $g_\mu:(g_0,g_1,g_2,g_3)=(-1,1,0,1)$ and $\sigma_0=\id_2$.
Both antilinear operators are invariant under $SL(2,\CC)$ operations 
on the qubit. 

One possible way of expressing the 3-tangle in terms of the above mentioned pair of
antilinear operators is
\beq\label{tau-3}
\tau_3[\psi]=\frac{1}{3}\bra{\psi*}\bullet\bra{\psi*}
\sigma_\mu\otimes \sigma_\nu\otimes \sigma_\lambda\otimes
\sigma^\mu\otimes \sigma^\nu\otimes \sigma^\lambda \ket{\psi}\bullet\ket{\psi} \; .
\eeq
This form has the minimal possible multi-linearity
and it is evidently permutation invariant.
Unfortunately, the convex roof construction presented in Refs.~\cite{Wootters98,Uhlmann00} 
for bilinear quantities can not in general be extended to higher multi-linearity.
This is very much related to the absence of a 
{\em tilde-orthogonal basis}~\cite{Wootters98}
that exists for bi-antilinear constructions (see ~\cite{Uhlmann00}). 

\begin{figure}[ht]
\includegraphics[width=.45\textwidth]{fig2-bloch.eps}
\includegraphics[width=.45\textwidth]{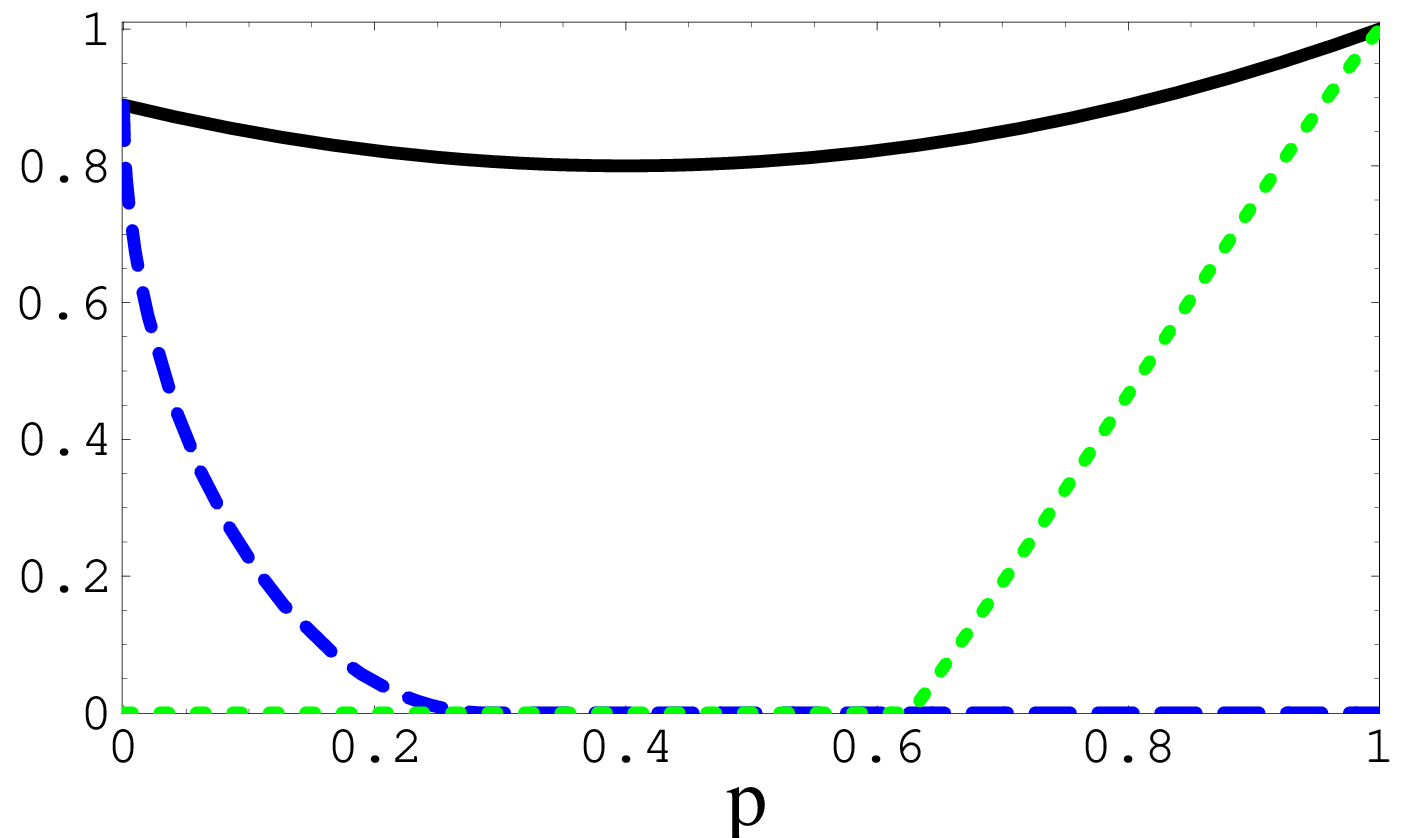}
\caption{{\em Left panel}: Bloch sphere for the two-dimensional
space spanned by the GHZ state and the W state. The simplex S0
contains all states with zero three-tangle. The leaves between p0
and p1 represent sets of constant three-tangle, and in the simplex
S1 the three-tangle is affine. The corners of the simplices constitute
the optimal decomposition.
{\em Right panel}: Plot of the convex roofs of the tangle (solid line), 
concurrence (dashed line) and 3-tangle (dotted line). There is an interval
with entangled states without concurrence or three-tangle.[From~\cite{LOSU}]}
\label{GHZ-W}
\end{figure}

A first success in direction of an understanding of convex roofs
for multipartite entanglement measures is the analytic
convex roof construction found for rank two-mixtures of translational 
invariant $GHZ$ and orthogonal $W$ states~\cite{LOSU}. 
The result is best expressed in the Bloch sphere of
both orthogonal states (left panel of Fig.~\ref{GHZ-W}).
The right panel of Fig.~\ref{GHZ-W} shows the convex roofs of the
three entanglement measures tangle(solid line), 
concurrence (dashed line) and 3-tangle (dotted line), which enter 
the monogamy relation of Coffman {\em et al.}.
The curious result is that an interval with vanishing concurrence
and 3-tangle appears, whereas the tangle is always positive.
On the background that only two classes of entanglement exist
for three qubits~\cite{Duer00}, this result might appear contradictory 
at first sight. That this is not the case,
is due to a subtlety of convex roof measures, when different
entanglement classes are considered
(here the concurrence - pairwise - and the 3-tangle).
The explanation is that the optimal decompositions leading to the
respective convex roofs, will in general not be compatible, in the sense that
both will not emerge from a single optimal decomposition of the whole 
density matrix.
The physical interpretation of this curious finding
is very simple: in the interval of
vanishing concurrence and 3-tangle, the density matrix can be built
from pure states that have either concurrence or 3-tangle (or both).

With the two independent local $SL(2,\CC)$-invariant operators (combs)
(\ref{Alin}) and (\ref{Alin-multi}) at hand, 
entanglement measures for genuine multipartite entanglement can now
be constructed for an arbitrary number of qubits.
The astonishing news from the four qubit case is that three 
inequivalent measures for SLOCC entanglement classes do exist~\cite{OS04}.
\begin{eqnarray}\label{4-filters}
{\cal F}^{(4)}_1 &=& 
                (\sigma_\mu\sigma_\nu\sigma_y\sigma_y)\bullet
                  (\sigma^\mu\sigma_y\sigma_\lambda\sigma_y)
        \bullet(\sigma_y\sigma^\nu\sigma^\lambda\sigma_y) \label{fourbit6lin}\\
{\cal F}^{(4)}_2 &=&{\mathfrak S}
                (\sigma_\mu\sigma_\nu\sigma_y\sigma_y)\bullet
                  (\sigma^\mu\sigma_y\sigma_\lambda\sigma_y)\bullet
   \nonumber\\
   &&\qquad\qquad \bullet        (\sigma_y\sigma^\nu\sigma_y\sigma_\tau) 
            \bullet
                (\sigma_y\sigma_y\sigma^\lambda\sigma^\tau)\label{fourbit8lin}\\
{\cal F}^{(4)}_3 &=&\bigfrac{1}{2}
                (\sigma_\mu\sigma_\nu\sigma_y\sigma_y)\bullet
                  (\sigma^\mu\sigma^\nu\sigma_y\sigma_y)
\bullet(\sigma_\rho\sigma_y\sigma_\tau\sigma_y) \bullet
         \nonumber\\ &&\qquad  \bullet    
          (\sigma^\rho\sigma_y\sigma^\tau\sigma_y)
 \bullet(\sigma_y\sigma_\lambda\sigma_\kappa\sigma_y) \bullet
                (\sigma_y\sigma^\lambda\sigma^\kappa\sigma_y)\label{fourbit12lin}
\end{eqnarray}
where ${\mathfrak S}$ indicates the symmetrization with respect to 
the permutation group $S_4$ on four qubits.
This leads to seven inequivalent representatives of SLOCC entanglement
classes for four qubits.\footnote{The seven representatives are characterized
by their different characteristics on the three measures 
(see Table~\ref{Table}): three each with one or two measures 
being non-zero plus one where all three measures are non-zero.}
These include the GHZ state~(\ref{2er}), 
the celebrated cluster state~(\ref{4er}) 
(three states connected by means of $S_4$ permutations) 
and a third state~(\ref{6er}), 
which in contrast to the former two has 
non-zero 4-qubit hyperdeterminant. 
\beqa
\label{2er}
    \ket{\Phi_2} &=& \frac{1}{\sqrt{2}}( \ket{0000}  +  \ket{1111})\\
\label{4er}
    \ket{\Phi_4} &=& \frac{1}{2}
    (\ket{1111}+\ket{1100}+\ket{0010}+\ket{0001})\\
\label{6er}
    \ket{\Phi_5} &=& \frac{1}{\sqrt{6}}
    (\sqrt{2}\ket{1111}+\ket{1000}+\ket{0100}+\ket{0010}+\ket{0001})
\eeqa
Furthermore, representatives of $q-1$ inequivalent 
non-zero SLOCC classes~\cite{Duer00,VerstraeteDM03} are known
for $q$ qubits~\cite{OS05}. 
Explicit measures for up to
six qubits and a prescription how general $q$-qubit entanglement 
measures are constructed have been given in Ref.~\cite{OS05}; their
characteristics on $q-1$ maximally entangled states demonstrates that
the states belong to different non-zero SLOCC classes (see table~\ref{Table}).
\begin{table}[h]\label{Table}
\begin{center}
\caption{Filter values for maximally entangled states; the length is the number
of Fock-elements in their canonical form. The 5- and 6-qubit entanglement 
measures are explicitly given in Ref.~\cite{OS05} together with the 
maximally entangled states they have been evaluated on here.}
\begin{tabular}{|c||c|c|c||c|c|c|c||c|c|}
\hline
length & $|{\cal F}^{(4)}_1|$ & $|{\cal F}^{(4)}_2|$ & 
$|{\cal F}^{(4)}_3|$ & $|{\cal F}^{(5)}_1|$ & $|{\cal F}^{(5)}_{2}|$ 
& $|{\cal F}^{(5)}_3|$ & $|{\cal F}^{(5)}_4|$ & $|{\cal F}^{(6)}_1|$ & $|{\cal F}^{(6)}_2|$ \\
\hline\hline
2 & 1               & 1& $\bigfrac{1}{2}$& 1                        & 1& 1                    & 
            $\bigfrac{1}{8}$ & 1 & 1\\
4 & 0               & \bigfrac{1}{3}& 1               & 0                        & 0& 0                    & 
                           1 & 0 & 0\\
5 & $\bigfrac{8}{9}$& 0& 0               & 0                        & 0& $\bigfrac{2^6}{3^5}$ & 
                           0 & 0 & 0 \\
6 &  X              & X& X               & $\bigfrac{3\sqrt{3}}{32}$& 0& 0                    & 
                           0 & 0 & 0\\
7 &  X              & X& X               & X                        & X& X                    & 
                           X & 0 &$\bigfrac{2^8}{5^5}$\\
\hline
\end{tabular} 
\end{center}
\end{table}

Summarizing, a method has been constructed that permits to construct
measures for genuine multipartite entanglement, i.e.
for the non-zero SLOCC class of entangled states.
Eventually necessary symmetrization with respect to the
corresponding symmetric group leads to generators of the ideal
of $SL*$-invariants~\cite{Dokovic} that vanish on arbitrary product states.
The generators of this ideal can also be constructed from the invariants
known from invariant theory~\cite{Luque02,Dokovic,Luque05}.
It would be highly desirable to bridge between both approaches.
For four qubits, this bridge has already been established
by the author in collaboration with D.~\v{Z}.~\DJo
to the extent that the three 4-qubit entanglement measures \mbox{have~been}
expressed in terms of three generators of that ideal but
constructed from the standard invariants from~\cite{Luque02,Dokovic}.
The results will be presented in a forthcoming publication\footnote{Note 
added after the habilitation procedure: this work is now available on the
arXiv:0804.1661}.

\section{Experimental access to genuine multipartite entanglement}
\label{Exp}

The invariants under local $SL(2,\CC)$ operations are most naturally expressed
in terms of expectation values of antilinear operators.
However, in the laboratory physical observables are measured,
which are linear operators.
So, unless the experiment is performed with quantum states with real 
coefficients only, a translation of the above multipartite entanglement 
measures into spin correlation functions will be necessary.
Such a one-to-one translation indeed exists. In order to see how the modulus of antilinear expectation values can be
transported into linear expectation values, we write
\beqa\nonumber
\bra{\psi}(\hat{O}\otimes &\dots&)\ket{\psi^*}\dots  \bra{\psi^*}
                           (\hat{O}\otimes \dots)\ket{\psi}\dots\\
&=&\bra{\psi}\bullet\bra{\psi^*}\bullet\dots
      \left[\hat{O}\bullet\hat{O}\bullet\dots\right]\otimes\dots 
                      \ket{\psi^*}\bullet\ket{\psi}\bullet\dots\\
&=&\bra{\psi}\bullet\bra{\psi^*}\bullet\dots\left[(\hat{O}\bullet\hat{O}\prod\P_{i,i`})\dots\right]\otimes\dots\ket{\psi}\bullet\ket{\psi^*}\bullet\dots 
\nonumber     
\eeqa
where $\P_{i,i`}$ is the permutation operator acting on the corresponding
different copies of the same qubit\footnote{$\P=\frac{1}{2}
\sigma_\mu\bullet\sigma_\mu$, using Einstein sum convention} 
and the product extends over all qubits and various copies such that
exclusively linear expectation values occur.
As a consequence, the linear operator corresponding to an antilinear operator ${\cal O}$ is then
\beq
{\mathfrak L}[{\cal O}]:=(\id\bullet{\mathfrak C})({\cal O}\bullet{\cal O})\P
\eeq
It is clear that the transformation in the opposite direction works the same way.
In particular we find
\beqa
{\mathfrak L}[\sigma_y] &=& 
M_{\mu\nu}\sigma_\mu\bullet\sigma_\nu \\
M_{\mu\nu} &=& \diag\{1,-1,-1,-1\} 
\eeqa
It is worth noticing that $M_{\mu\nu}$ is precisely the Minkowski metric.
\beqa
{\mathfrak L}[\sigma_\mu\bullet\sigma^\mu] &=&{\mathfrak G}_{\kappa\lambda\mu\nu}
\sigma_\kappa\bullet\sigma_\lambda\bullet \sigma_\mu\bullet\sigma_\nu \\
\!\!\!\!\!{\mathfrak G}_{\kappa\lambda\mu\nu} &=& 
2M_{\kappa\lambda}M_{\mu\nu}-M_{\kappa\mu}M_{\lambda\nu}+2M_{\kappa\nu}M_{\lambda\mu}
\eeqa
This is a straight forward prescription for expressing the genuine multipartite entanglement
measures
presented in Ref.~\cite{OS04,OS05} in terms of expectation values of 
linear operators.
The details of this transcription including an analysis of
weakly mixed states will be presented in a forthcoming publication.

${}$
\newpage
${}$
\newpage

\chapter{Conclusions \& Outlook}
In conclusion, quantum entanglement is a fascinating phenomenon that
naturally occurs in the eigenstates of interacting Hamiltonians.
It is certainly the interpretation of this entanglement as a resource
for quantum information processing that nowadays is mainly responsible 
for its growing relevance and consideration in e.g. condensed matter physics.
Nevertheless did measures for certain classes of entanglement
also prove to be useful tools for the indication of quantum phase transitions
if the corresponding measure (e.g. some entanglement entropy) incorporates
the relevant order parameter. Though there is only a single class-selective
entanglement measure we can compute for mixed states, namely the concurrence
for pairwise entanglement of qubits, the Coffman-Kundu-Wootters relation
makes the local von Neumann entropy to an implicit 
indicator of multipartite entanglement.
The analysis of ground state entanglement of many model Hamiltonians
gives strong support to the conclusion
that multipartite entanglement becomes dominant in quantum Hamiltonians
close to their critical point. But it is also rather multipartite than only pairwise entanglement,
that is created by Hamiltonian evolution from disentangled initial states.
A deeper analysis is needed here in particular in the direction of more conclusive measures
rather than analyzing more and more different Hamiltonians with always the same 
canonical set of measures. 
On the other hand, since the presence of multipartite entanglement seems to be the generic case
rather than the exception,
it might be the weakly or even disentangled states, as for the factorizing field 
in spin chains, which are the precursors of drastic changes and complex phenomena.
In any case must the understanding of entanglement and its subdivision into distinct SLOCC classes 
be improved substantially before a possible connection between entanglement and quantum phase 
transitions can be established. A step into this direction is to realize that the relevant
symmetry group for entanglement classification is the local $SL(2,\CC)$, including the
permutation group of the qubits, and that entanglement monotones for this invariance group
can be built from local invariant and antilinear 
operators with zero expectation values on all the local Hilbert space - combs. 
Two such operators have been found recently for qubits. 
They permit the construction of a complete set of measures for
genuine multipartite entanglement provably up to 4 qubit systems. 
The main achievement lies in that the combs admit a straight forward construction
of the ideal of $SL(2,\CC)^{\otimes N}$-invariants whose elements vanish on all product states
and for arbitrary number of qubits $N$;
this is a problem which already for five qubits creates severe difficulties for standard methods
from invariant theory, where it seems hopeless to handle them for systems larger than this.
A one-to-one correspondence of the antilinear invariants to correlation function finally
even gives a prescription for the laboratory in order to extract these quantities  from
spin correlation functions.
Yet, this is rather the very beginning of research in this direction, since there is little understanding
of even very elementary questions.
Although a set of maximally entangled states is given for an arbitrary number of qubits,
the ambiguity of such a set is an issue: it is not clear from the beginning, whether a complete
set of maximally entangled states can possibly be constructed from them as elements of
kind of a basis for such states. A characterization of maximally entangled states 
might be thinkable in terms of the analogue of a {\em tilde-orthogonal basis} 
(see~\cite{Wootters98}).
Completeness of the ideal is another big issue which would have an important impact also 
in the field of invariant theory.
A further major challenge consists in the extension of the pure state measures to mixed states, which
at the end is necessary in order to express the entanglement measure in terms of correlation
functions of a practically infinitely large condensed matter system.
Also here an important step forward has been done on the mixture of three qubit 
GHZ and W states, but the ultimate goal would be an analytic procedure for
the convex roof extension along the lines of the concurrence.
Already now the accumulated knowledge about optimal decompositions is of substantial
help for numerical procedures and leads to nontrivial lower bounds for multipartite entanglement
in mixed states, but the major part of the puzzle pieces is yet to be found.
A consequent extension of the method of combs would consist in their identification
for local Hilbert space dimension larger than two. 
Here the problem is more complex already at the starting point,
because there are no bi-antilinear $SL(2S+1,\CC)$ invariant combs for spin $S>\frac{1}{2}$.
Last but not least we mention that once, genuine multipartite entanglement measures are
known, they should be used for testing and analyzing alternative approaches to the detection of entanglement. Witnesses are one relevant example of a complementary approach, which is motivated 
from the question for separability of a mixed quantum state rather than from the classification
of entanglement. It might be just their complementarity bearing the key towards many open questions
in the theory of entanglement.

\providecommand{\bysame}{\leavevmode\hbox to3em{\hrulefill}\thinspace}
\providecommand{\MR}{\relax\ifhmode\unskip\space\fi MR }
\providecommand{\MRhref}[2]{%
  \href{http://www.ams.org/mathscinet-getitem?mr=#1}{#2}
}
\providecommand{\href}[2]{#2}

\end{document}